\documentclass{pasj01}
\draft
\usepackage{xcolor,ulem}
\usepackage{longtable}
\usepackage{lscape}
\usepackage{dcolumn}
\usepackage{comment}
\newcolumntype{d}[1]{D{.}{.}{#1}}
\newcommand{\fn}[1]{{\color{red}{#1}}}

\Received{$\langle$reception date$\rangle$}
\Accepted{$\langle$acception date$\rangle$}
\Published{$\langle$publication date$\rangle$}

\begin{document}

\title{
The C$^{18}$O core mass function toward Orion A: Single-dish observations
}
\author{Hideaki \textsc{Takemura}\altaffilmark{1,2},
Fumitaka \textsc{Nakamura}\altaffilmark{1,2,3},
Shun \textsc{Ishii}\altaffilmark{1,2},
Yoshito \textsc{Shimajiri}\altaffilmark{2},
Patricio \textsc{Sanhueza}\altaffilmark{1, 2},
Takashi \textsc{Tsukagoshi}\altaffilmark{2},
Ryohei \textsc{Kawabe}\altaffilmark{1,2},
Tomoya \textsc{Hirota}\altaffilmark{1,2},
Akimasa \textsc{Kataoka}\altaffilmark{1,2}
}%

\altaffiltext{1}{The Graduate University for Advanced Studies
(SOKENDAI), 2-21-1 Osawa, Mitaka, Tokyo 181-0015, Japan}
\altaffiltext{2}{National Astronomical Observatory of Japan, 2-21-1 Osawa, Mitaka, Tokyo 181-8588, Japan}
\altaffiltext{3}{Department of Astronomy, The University of Tokyo, Hongo, Tokyo 113-0033, Japan}


\maketitle

\begin{abstract}
We have performed an unbiased dense core survey toward the Orion A Giant Molecular Cloud
in the C$^{18}$O ($J$ = 1--0) emission line taken with the Nobeyama Radio Observatory (NRO) 45-m telescope.
The effective angular resolution of the map is 26\arcsec, which corresponds to $\sim$ 0.05 pc at a distance of 414 pc.
By using the \textit{Herschel}--\textit{Planck} H$_2$ column density map, we calculate the C$^{18}$O fractional abundance and find that it is roughly constant over the column density range of
$\lesssim$ 5 $\times$ 10$^{22}$ cm$^{-3}$, although a trend of C$^{18}$O depletion is determined toward higher column density.
Therefore, C$^{18}$O intensity can follow the cloud structure reasonably well.
The mean C$^{18}$O abundance in Orion A is estimated to be 5.7$\times$10$^{-7}$, which is about 3 times larger than the fiducial value.
We identified 746 C$^{18}$O cores with \texttt{astrodendro} and
classified 709 cores as starless cores.
We compute the core masses by decomposing the \textit{Herschel}--\textit{Planck} dust column density using the relative proportions of the C$^{18}$O integrated intensities of line-of-sight components.
Applying this procedure, we attempt to remove
the contribution of the background emission, i.e., the ambient gas outside the cores.
Then, we derived mass function for starless cores and found that it resembles the stellar initial mass function (IMF).
The CMF for starless cores, $dN/dM$, is fitted with a power-law relation of $M^\alpha$ with a power index of $\alpha = -$2.25$\pm$ 0.16 at the high-mass slope ($\gtrsim$ 0.44 $M_\odot$).
We also found that the ratio of each core mass to the total mass integrated along the line of sight is significantly large.
Therefore, in the previous studies, the core masses derived from the dust image are likely to be overestimated at least by a factor of a few.
Accordingly, such previous studies may underestimate the star formation efficiency of individual cores.
\end{abstract}

\section{Introduction}
\label{sec:intro}
Stars are formed in dense cores embedded in molecular clouds.
Therefore, it is important to understand how dense cores form from parental molecular clouds.
Previous studies of nearby low-mass star-forming regions have suggested that a typical dense core has its mass of 1--10 $M_\odot$, size of 0.01--0.1 pc, and density of $10^{4-5}$ cm$^{-3}$ \citep{bergin07}.
Since the evolutionary processes of stars are partly determined by their masses at birth,
it is important to reveal the origin of the mass distribution of dense cores in molecular clouds.
Thus, the mass functions of dense cores (Core Mass Functions, CMFs) are expected to imprint some information on their formation and evolution processes.
\citet{salpeter55} derived the stellar initial mass function (IMF) for nearby stars and he discovered that the IMF has a power-law shape
of $dN/dM \propto M^{-2.35}$ at the high-mass end
($\gtrsim$ 1 $M_\odot$).

Both the CMF and IMF are often characterized by their slopes at the high-mass ends ($\gtrsim$ 1 $M_\odot$) and the turnover masses at the low-mass part of the distributions.
Many previous studies toward nearby star-forming regions have suggested that the CMF has a Salpeter-like slope at the high-mass end
and resembles the IMF \citep{motte98,alves07,konyves10}.
This resemblance of the CMF and IMF appears to suggest that the identified cores are the immediate precursors of stars.
One fundamental difference, however, between the CMF and the IMF is the turnover mass, i.e., the peak at the low-mass part.
Observed CMFs usually have larger turnover masses than observed IMFs \citep{alves07, nutter07, konyves10}.
The previous studies proposed that this difference is consistent with the idea that about half of each core mass is blown out by stellar feedback such as protostellar outflows and stellar winds \citep{matzner00,alves07}.
However, recent observations of CMF in the W43-MM1 high-mass star-forming region revealed a CMF with a shallower slope. This might indicate that the CMFs evolve with time \citep{motte18}.
On the other hand, \citet{bontemps01} revealed that the CMF in $\rho$ Oph has a turnover mass similar to that of the mass function of Class II objects in this region. This indicates that almost all the core masses should go into the stars formed.
\citet{ikeda07} discussed the effect of confusion among cores on CMF in Orion A, especially to the low-mass part.
They claimed that there is no turnover in confusion-corrected CMF and the observed turnover is not made by any physical processes.

Most of the previous studies on the CMFs have used the dust continuum emission \citep{motte98, alves07, nutter07, konyves10, motte18, liu18, kong19, sanhueza19}. In general, the dust emission is optically thin on the cloud scale and thus it is an excellent tracer of the cloud column density since the spatial variation of the dust-to-gas mass ratio is believed to be small.
However, a core is a dense structure embedded in the parent cloud. Thus, it seems to be very difficult to accurately estimate the core masses from the 2D dust emission maps.
In other words, it is crucial to estimate the core masses by removing the ambient gas outside the core.
In this paper, we call gas components that are not associated with the core along the line of sight as ambient gas.
In this sense, the previous studies of the CMFs based on the dust emission and extinction maps tend to overestimate the core masses.

In this paper, toward a full understanding of the relationship between the CMF and IMF, we attempt to identify the cores in the Orion A Giant Molecular Cloud using 3D position-position-velocity data of a wide-field C18O mapping and estimate the core masses using the \textit{Herschel}--\textit{Planck} column density map \citep{lombardi14} by removing the contribution of the ambient gas.
It is worth noting that previous studies have demonstrated that real structures in molecular clouds are reasonably related to the structures identified in the position-position-velocity data \citep{williams94,rosolowsky08,goodman09}.

Orion A is the most studied giant molecular cloud \citep{bally87, genzel89, hillenbrand97, ikeda07, shimajiri11, nakamura12, tatematsu16, hacar17}. The distance to the cloud is derived to be 414 pc based on the VLBI observations \citep{menten07}.
Thus, it is one of the nearest GMCs.
Orion A also contains the nearest ongoing high-mass star formation site in Orion Nebula Cluster. Near-infrared observations also indicate that younger populations are more abundant, i.e., star formation is accelerated \citep{palla99}.
Recently, we have obtained wide-field maps of this region in $^{12}$CO ($J$=1--0), $^{13}$CO ($J$=1--0), and C$^{18}$O ($J$=1--0) with the Nobeyama 45-m telescope \citep{feddersen18, nakamura19,ishii19,tanabe19}.
Among the three CO isotopologues, the C$^{18}$O emission is expected to trace denser structures with densities of $\sim 10^4$ cm$^{-3}$.
Recent studies by \citet{pety17} and \citet{gratier20} also demonstrated that the C$^{18}$O emission reasonably traces the high-column density molecular gas with $N_{\rm H_2} \sim 10^{22}$ cm$^{-3}$.
The wide-field H$_2$ column density map obtained from the \textit{Herschel} observations, which is derived from the dust emission, is also available \citep{lombardi14}.
Thus, Orion A is one of the suitable star-forming clouds to assess the contribution of the ambient gas mass to the core mass estimate.

This paper is organized as follows.
In section \ref{sec:obs}, we present the details of the observations and data. In section \ref{sec:global}, we describe the global distribution of the dense molecular gas with the C$^{18}$O data. Then, we show the results of the core identification in section \ref{sec:core_id}. Here, we adopt Dendrogram and define a leaf as a core, following the previous studies
In section \ref{sec:discussion}, we present the CMF derived from the C$^{18}$O data. We also compare our results with previous studies of the CMF in Orion A.
In section \ref{sec:dust_c18o}, we compare the CMF derived from the C$^{18}$O data and dust continuum emission \citep{nutter07}.
Finally, we briefly summarize our results in section \ref{sec:summary}.

\section{Observations and data}
\label{sec:obs}

\subsection{C$^{18}$O ($J$=1--0) }

\begin{figure}[htbp]
 \begin{center}
  \includegraphics[scale=0.3,bb=0 0 1464 1754]{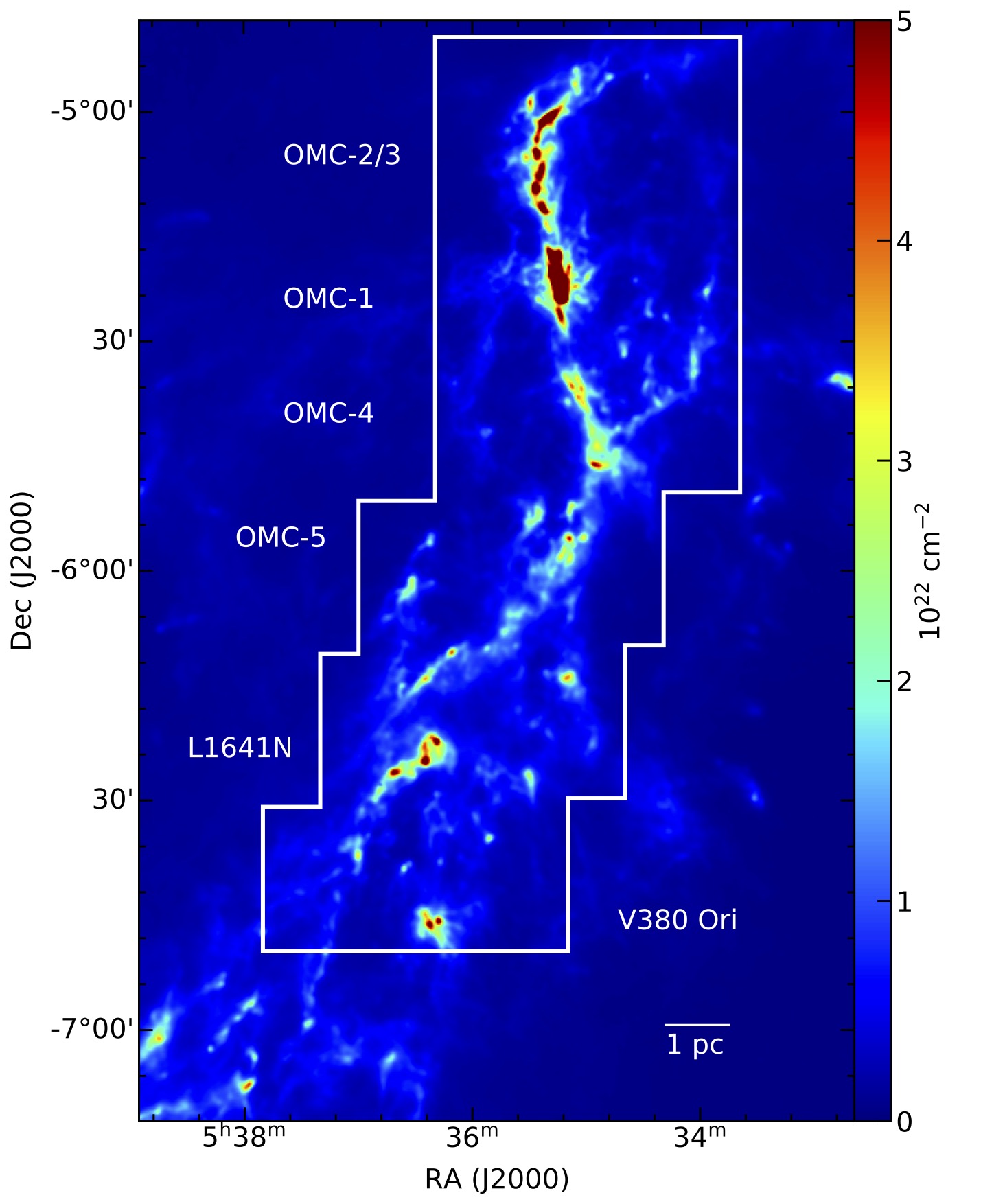}
 \end{center}
\caption{The C$^{18}$O ($J$=1--0) observation area overlaid on the \textit{Herschel}--\textit{Planck} H$_2$ map \citep{lombardi14}.
The observed area is indicated with the white solid line.
Our map covers a wide area of 1 $\times$ 2 square degree which is from OMC-1/2/3/4 to L1641N, and V380 Ori.}
\label{fig:obsarea}
\end{figure}

We carried out On-The-Fly (OTF) mapping observations of C$^{18}$O ($J$ = 1--0, 109.782182 GHz) toward Orion A
using the FOREST \citep{minamidani16} receivers mounted on the NRO 45-m telescope.
Our map covers 1 $\times$ 2 square degree area which contains OMC-1/2/3/4, L1641N, and V380 Ori.
Figure \ref{fig:obsarea} shows the observed area of the C$^{18}$O emission overlaid on the H$_2$ column density map obtained with the procedure described in the next subsection.

The details of the FOREST observations are described in \citet{nakamura19}.
In brief, the observations were done in the period from 2016 March to 2017 March.
The telescope beam size (HPBW) is $\sim$15\arcsec \ at 110 GHz and the typical pointing accuracy was 3\arcsec.
The effective angular resolution of the data product is $\sim$21\arcsec\ (FWHM) after convolving by a spheroidal function with a spatial grid size of 7.\arcsec5.
In this paper, the data was additionally smoothed to reduce the noise level of the map.
The C$^{18}$O map has an effective resolution of $\sim$ 26.\arcsec4 (FWHM), corresponding to $\sim$ 0.05 pc at a distance of 414 pc, and a velocity resolution of $\sim$ 0.1 km s$^{-1}$.
The typical rms noise level of the C18O data is 1$\sigma$ = 0.33 K in the unit of T$_{\rm MB}$.
It is worth noting that the original map provides a higher angular resolution of $\sim 21$\arcsec, but we smoothed the data to a lower angular resolution to reduce the noise level.
We used the data taken only with the FOREST receiver in this paper.
\citet{ikeda09} derived CMF in a square degree area of Orion A from C$^{18}$O ($J$=1--0) observations with the comparable resolution taken by the same telescope. We compare the result of our analysis with \citep{ikeda09} in section \ref{sec:cmf}.
Our observations cover a larger area of 1$\times$2$\degree$ than that of \citet{ikeda09}, who conducted the mapping observations toward the Orion Nebula Cluster region in a full-beam sampling mode with a previous receiver, BEARS, and the noise level of our map (0.33 K) is better than theirs (0.45K). The velocity resolution of $\sim 0.1$ km s$^{-1}$ is almost similar to each other.
The integrated intensity map of the C$^{18}$O ($J$ = 1--0) emission is shown in figure \ref{fig:mom012} (a).

\subsection{H$_2$ column density data}

We also used the \textit{Herschel}--\textit{Planck} H$_2$ column density map
\citep{lombardi14,stutz15} to calculate the core masses.
The original Herschel data have an effective angular resolutions of 18 and 36\arcsec \ at the 250 and 500 $\mu$m, respectively.
The temperature map with 36\arcsec \ and 250 $\mu$m continuum emission map with 18\arcsec \ are used
to make the column density map. See \citet{stutz15} for the actual procedure.

Then, we smoothed the 18\arcsec \ data to match the C$^{18}$O effective angular resolution of 26\arcsec.4, with a grid of 7\arcsec. 5.

\subsection{Catalog of young stellar objects (YSOs)}

In order to distinguish starless cores from protostellar cores,
we used a catalog of YSOs from the Herschel Orion Protostar (HOP) Survey \citep{furlan16}.
This catalog contains 36 Class 0 sources and 38 Class I sources in our observed area.
In this paper, we use the 74 Class 0 and Class I sources as HOPS YSOs.

\section{Global distribution of the C$^{18}$O emission}
\label{sec:global}

\begin{figure}[htbp]
 \begin{center}
  \includegraphics[scale=0.27,bb=0 0 1707 807]{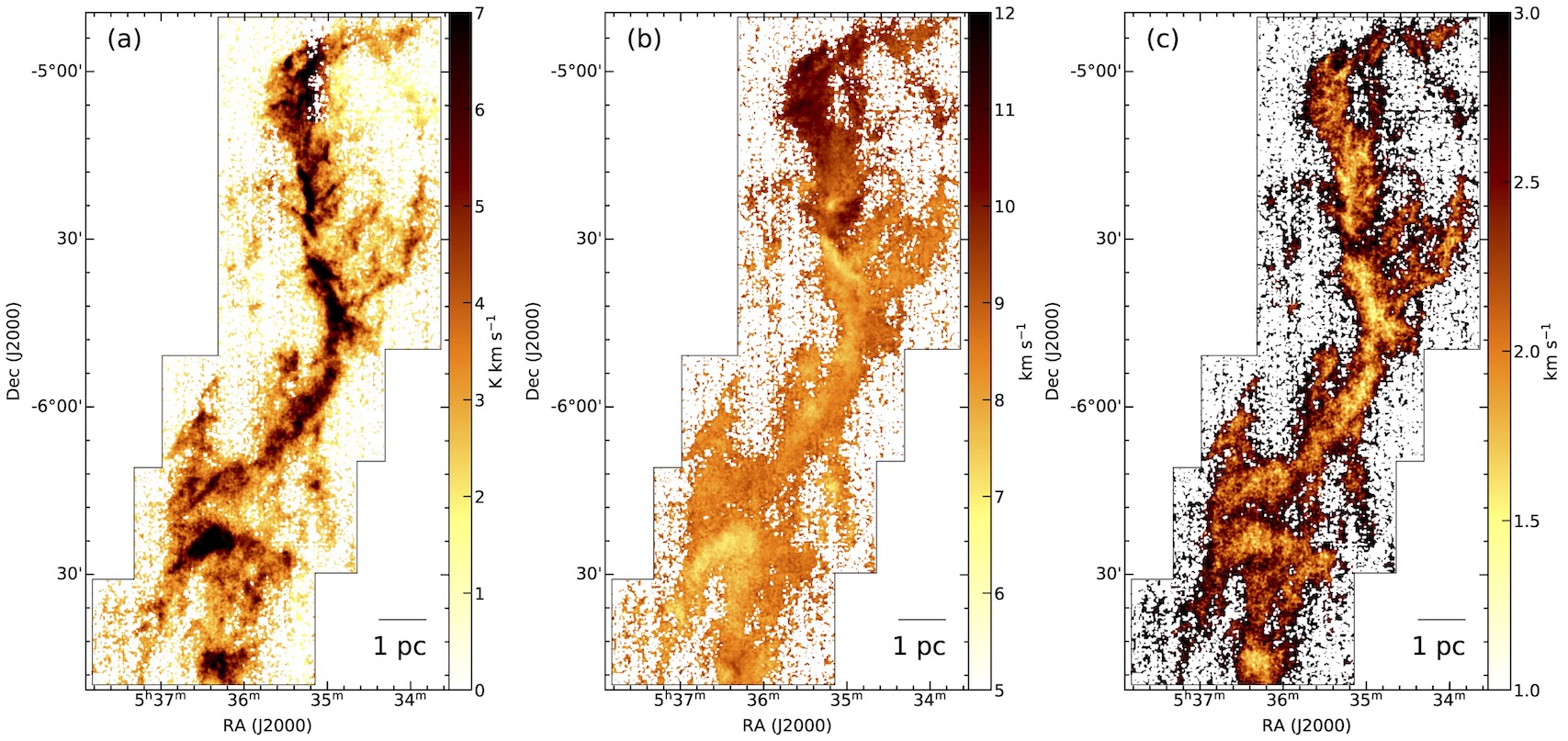}
 \end{center}
\caption{(a) Integrated intensity map (moment-0 map) of the C$^{18}$O ($J$=1--0) emission,
(b) intensity weighted coordinate (moment-1 map) and (c) intensity weighted dispersion of the coordinate (moment-2 map), respectively.
The integration was done in the velocity range from $\sim$ 3 km s$^{-1}$ to $\sim$ 15 km s$^{-1}$ for the pixels whose integrated intensity is larger than 15 local rams noise.
We made these maps with the CASA software.
The observed area is indicated with the black solid line in each panel.}
\label{fig:mom012}
\end{figure}

Figure \ref{fig:mom012} (b) and (c) show the maps of the intensity weighted velocity and the velocity dispersion of Orion A.
The C$^{18}$O emission well follows the long filamentary structure, which includes the OMC-1/2/3/4 regions, called
the Intengral-shaped Filament (ISF).
Also, there is a velocity gradient along the ISF.
The northern region has a velocity of $\sim$ 12 km s$^{-1}$ and the southern region such as L1641N has a velocity of $\sim$ 5 km s$^{-1}$.
The moment-1 map [figure \ref{fig:mom012} (b)] also shows a sudden jump in the velocity from $\sim$ 9 km s$^{-1}$ to $\sim$ 5 km s$^{-1}$ around the L1641N region, and V380 Ori region has a velocity of $\sim$ 9 km s$^{-1}$. In other words, two components with different velocities appear to overlap along the line-of-sight.
\citet{nakamura12} suggested that these components may be colliding (see also Lim et al. 2020). In fact, some faint extended emission with intermediate velocities appears to connect with the two components in the position-velocity diagrams (see figure 8 in \citet{nakamura12}).
The OMC-1 region has a very complicated velocity structure, which
presumably is affected by stellar feedback or global gravitational contraction along the ISF \citep{hacar17, ishii19}
and cloud-cloud collision \citep{fukui18}.
The moment-2 map [figure \ref{fig:mom012} (c)] indicates that
the velocity dispersion tends to be larger toward brighter regions inside the ISF.

\begin{figure}[htbp]
 \begin{center}
  \includegraphics[scale=0.3,bb=0 0 1246 1579]{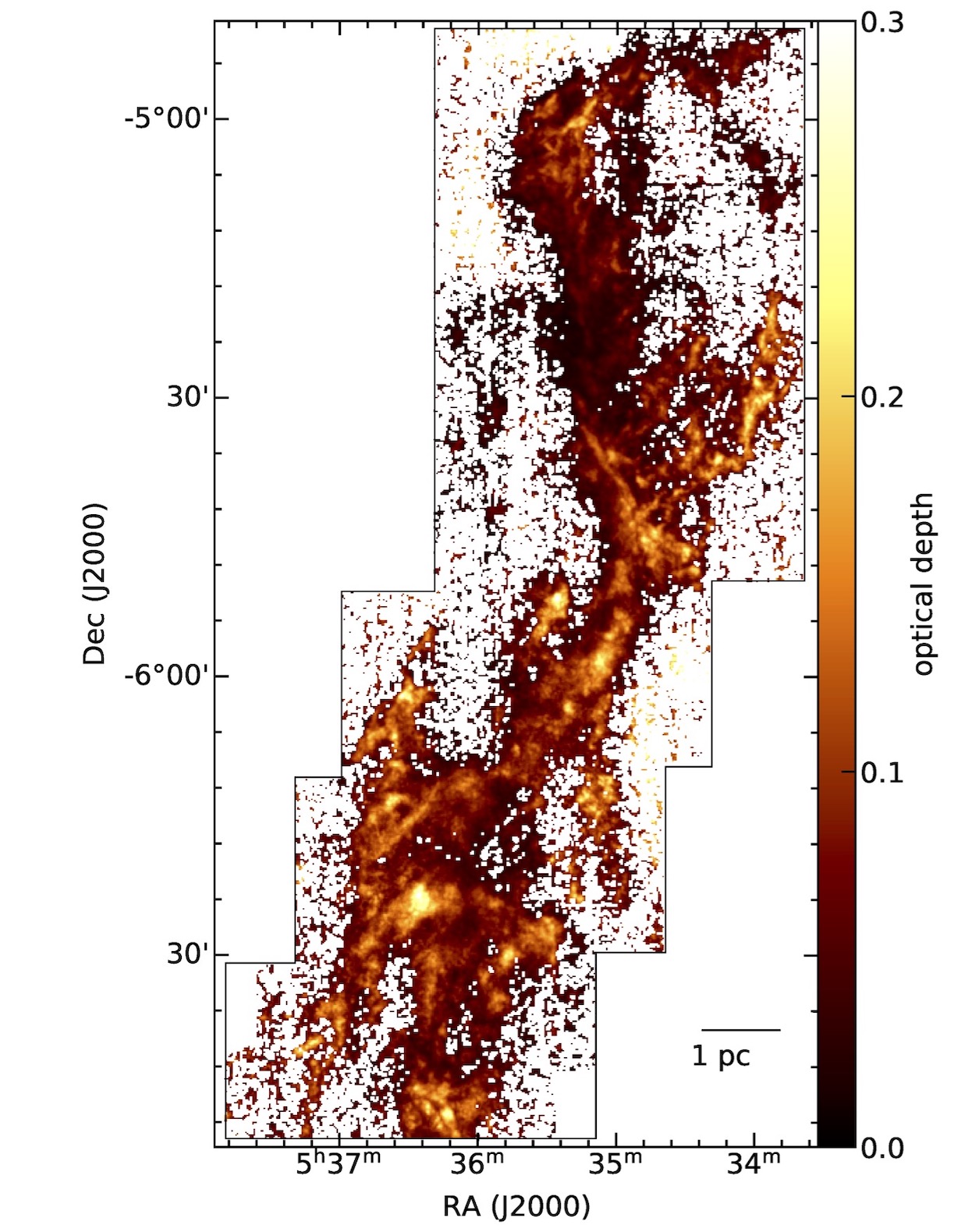}
 \end{center}
\caption{
The optical depth of the C$^{18}$O ($J$ = 1--0) emission in Orion A.
We calculated the optical depth for the same pixels of figure \ref{fig:mom012} by using the temperature derived from $^{12}$CO ($J$=1--0) observations.}
\label{fig:optical depth}
\end{figure}

Figure \ref{fig:optical depth} shows the optical depth of C$^{18}$O in Orion A.
Here, we adopted Equation (6) in \citet{nakamura19} with the excitation temperature derived with their Equation (3).
The maximum optical depth is about 0.2 and thus we assume that the C$^{18}$O emission is optically thin in the entire area.

\section{Core identification}
\label{sec:core_id}

There are at least two main approaches to identify the dense cores in molecular clouds. One is to use dust emission or extinction maps (e.g., \cite{motte98}; \cite{alves07}; \cite{nutter07}; \cite{cheng18}; \cite{sanhueza19}). 
The other is to use molecular emission line data (e.g., \cite{ikeda07}; \cite{ikeda09}; \cite{maruta10}). 
The dust emission is considered to be an excellent tracer of the column density of molecular clouds or interstellar medium (ISM).
However, the dust cores, which have a three-dimensional structure in nature, are identified on the basis of the two-dimensional position-position maps.
It is difficult to distinguish the structures overlapped along the line of sight.

The structure identification using molecular line emission allows to distinguish overlapping structures along the line of sight by taking advantage of
the three-dimensional position-position-velocity (p-p-v) data. This method assumes that the coherent structures in p-p-v space are closely related to actual structures in position-position-position (p-p-p) space.
However, the fractional abundances of the interstellar molecules relative to the hydrogen gas often vary even in single clouds \citep{savva03}. Thus, the mass estimation would be significantly influenced by the spatial variation of the fractional abundances.
In this sense, using these two tracers is complimentary for the structure identification in molecular clouds.

In this paper, we use the advantages of the two tracers for core identification.
We first identify the dense cores using our C$^{18}$O ($J$ = 1--0) data cube with a velocity resolution of 0.1 km s$^{-1}$.
Then, we use the \textit{Herschel}--\textit{Planck} H$_2$ column density map to estimate the masses of the cores identified from the C$^{18}$O ($J$ = 1--0) data.
To evaluate the masses of the cores, we attempt to remove the contribution of the ambient gas to the core mass estimation using the C$^{18}$O ($J$ = 1--0) data.
As shown above, the C$^{18}$O emission is optically-thin. However, the C$^{18}$O molecule tends to be sometimes depleted in cold dense regions with $T \lesssim 20$ K \citep{caselli99}.
In the presence of strong far UV radiation, selective dissociation changes the fractional abundances of CO and its isotopologue \citep{lada94, shimajiri14,lin16,ishii19}. These effects tend to destroy the linear relationship between the intensities of the dust emission and the optically-thin C$^{18}$O emission.

\subsection{Dendrogram and dense cores}

We applied astrodendro ver. 0.2.0
\footnote{https://dendrograms.readthedocs.io/en/stable/} to the three-dimensional (p-p-v) C$^{18}$O data.
The algorithm searches the hierarchical structures in the two-dimensional data (position-position)
or three-dimensional data for given parameters.
The astrodendro has three input parameters, (1) \texttt{min\_value}, (2) \texttt{min\_delta}, and (3) \texttt{min\_npix}.
It keeps searching the local peaks from the highest value of the data to the lowest value specified by \texttt{min\_value} in the interval of \texttt{min\_delta}.
The local peak is called \texttt{leaf} in the algorithm.
In addition, the number of pixels contained in a leaf should be greater than \texttt{min\_npix}.

Two leaves are merged into one structure called
\texttt{branch}
when the bottoms of both leaves are larger than \texttt{min\_value}.
After that, two branches are merged into a new branch or a \texttt{trunk} which is the lowest structure.
Then, the result of the identification is depicted with a tree-like diagram.
A more detailed description and discussion about the algorithm are given in \citet{rosolowsky08}.
See also \citet{goodman09} for the application to the 3D molecular line data.
In this study, we adopted \texttt{min\_delta} = 2$\sigma$, \texttt{min\_value} = 2$\sigma$ and \texttt{min\_npix} = 30 which corresponds to the volume of 1 beam times 3 channels for our structure identification.
We define a leaf as a dense core of the molecular cloud.
We note that this core definition is reasonably verified by the synthetic observations of the turbulent cloud simulations \citep{burkhart13,beaumont13}, although the structures identified in the real p-p-p space do not perfectly match with those identified in the p-p-p space. In addition, the structure identification in the p-p-p space also depends on cloud environments.

Since the observations were carried out in various atmospheric conditions for a couple of years, the
rms noise level of the map is not uniform over the observed area. The value of \texttt{min\_value} does not always correspond to twice the noise level. To minimize the effect of the non-uniform
noise levels in the core identification, we imposed the following three conditions for identified leaves with the parameters mentioned above. Condition (1): the peak intensity of the identified leaf should be four times larger than the local rms noise level at the corresponding spatial position.
Condition (2):
More than three successive channels ($\sim$ 0.3 km s$^{-1}$) should contain more than 9 pixels for each channel.
This threshold pixel number of 9 is equivalent to one effective angular resolution of the map ($\approx$ 26.\arcsec4).
Condition (3):
an identified core should not contain any pixels located at the boundaries of the observation area.
In total, we identified 746 cores.

Then, we classified the identified cores into two groups using the HOPS catalog, starless and protostellar cores.
If an identified core overlaps spatially with at least one HOP object on the plane of the sky, we classified it as a protostellar core. Here we classified an identified core as a protostellar core when it contains more than one HOP object within the projected area onto the plane of the sky.
A core without the overlapped HOP objects is categorized as a starless core.

\begin{figure}[htbp]
 \begin{center}
  \includegraphics[scale=0.3,bb=0 0 1384 1754]{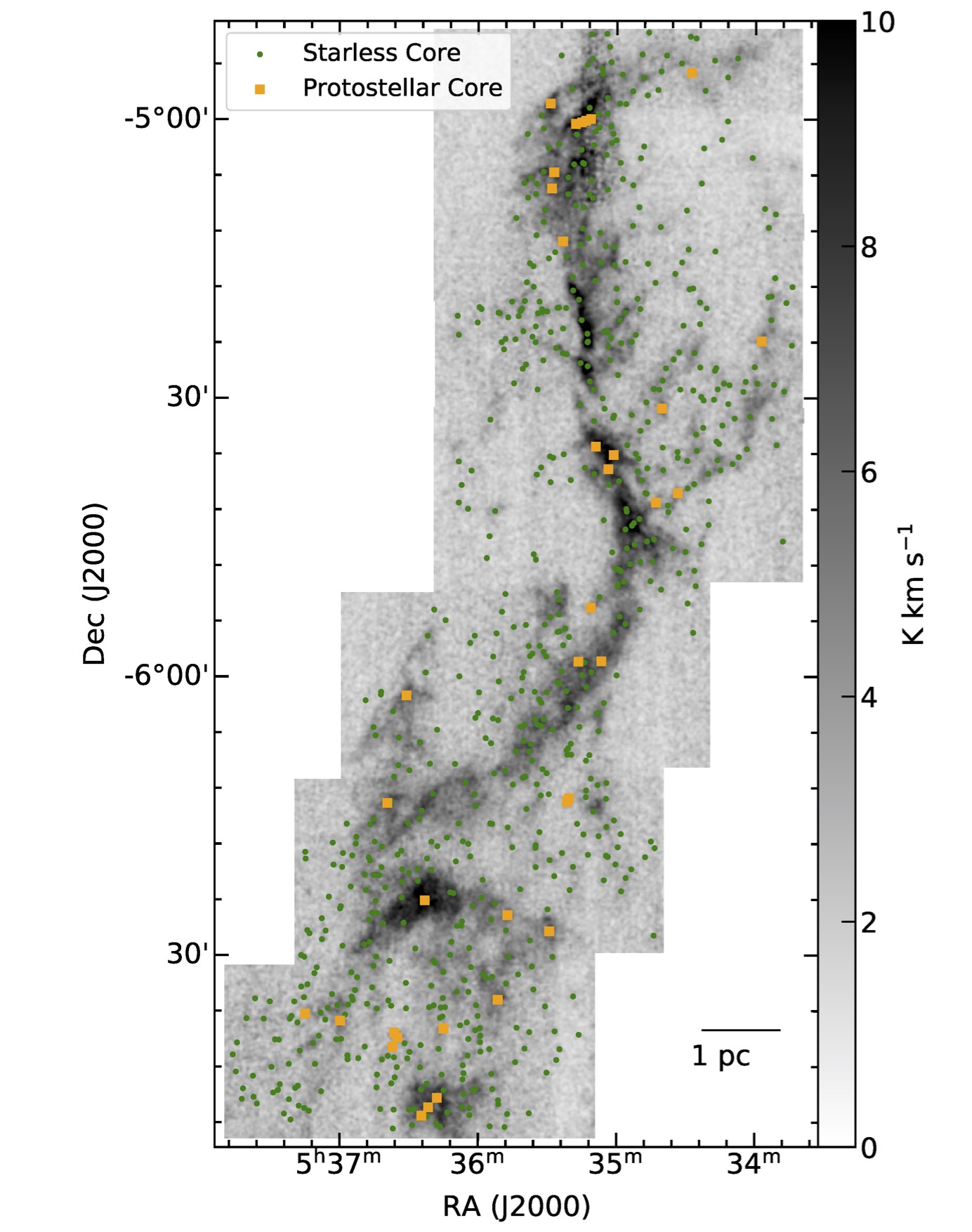}
 \end{center}
\caption{The identified starless cores (green circles) and protostellar cores (orange squares)
are plotted on the integrated intensity map of the C$^{18}$O emission.
}
\label{fig:mom0_sl_ps}
\end{figure}

As a result, we identified 709 starless cores and 37 protostellar cores (table \ref{tab:oriona_core_id}). Figure \ref{fig:mom0_sl_ps} shows the result of the core identification.
The starless cores and the protostellar cores are plotted onto the integrated intensity map
of the C$^{18}$O emission and most cores are distributed along ISF.
Among the 74 HOPS Class 0/I objects, 34 ($\sim$46 \%) objects were not identified as our protostellar cores. However, most of the HOPS Class 0/I objects are associated with local C$^{18}$O peaks and they are excluded from our samples by the additional thresholds of core identification. To identify these protostellar cores, we need data with better sensitivities.

\subsection{Derivation of the core physical quantities}

\subsubsection{Core radius, aspect ratio, and velocity dispersion}

\begin{figure}[htbp]
 \begin{center}
  \includegraphics[scale=0.2,bb=0 0 1749 2660]{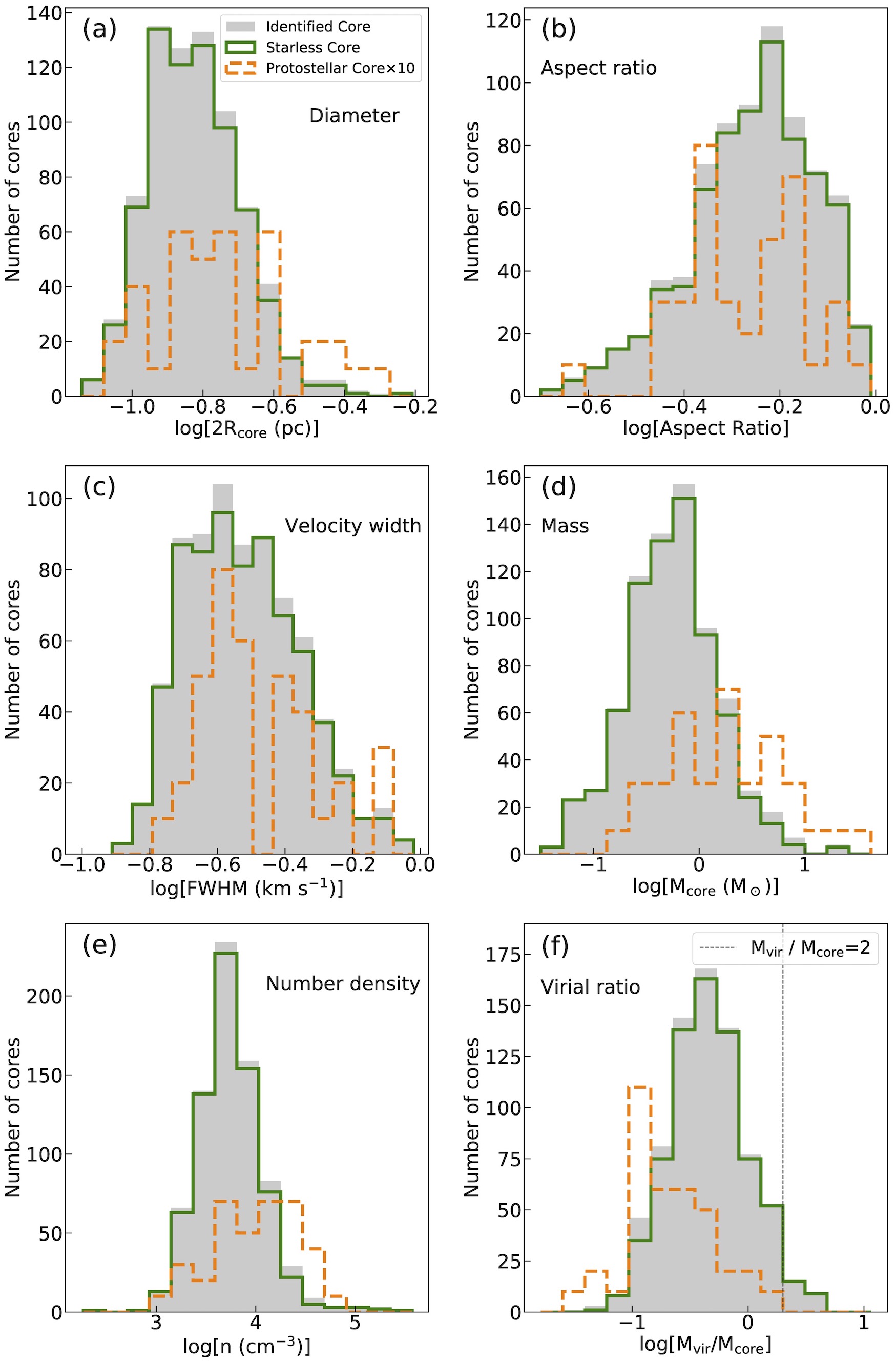}
 \end{center}
\caption{The histograms of (a) the diameter (b) the aspect ratio, (c) the FWHM velocity width, (d) the core mass, (e) the number density and
(f) the virial ratio of C$^{18}$O cores calculated from C$^{18}$O observation, respectively.
The histogram for all identified cores is shown in grey.
The green and orange histograms are for starless cores and protostellar cores.
The vertical line in (d) shows $M_\mathrm{core}/M_\mathrm{vir}==2$.}
\label{fig:histo_sl_ps}
\end{figure}

We define the physical quantities of the identified cores as follows.
The positions and the line-of-sight velocity of a core are determined by the mean positions of the structure identified and the intensity-weighted first-moment velocity, respectively.
The core radius is defined as
\begin{equation}
    R_{\mathrm{core}}=\left(\frac{A}{\pi}\right)^{1/2}  \ ,
\end{equation}
where $A$ is the exact area of the core projected onto the plane of the sky.
The aspect ratio of the core is calculated as the ratio of the major and minor axes.
The major and minor axes in half width at half maximum (HWHM) are computed from the intensity-weighted second moment
in the direction of greatest elongation and perpendicular to the major axis, respectively,
in the plane of the sky. The position angle of the core is determined
counter-clockwise from the +R.A. axis.
The velocity width in in full width at half maximum (FWHM), $dV_{\mathrm{core}}$ is obtained by multiplying
the intensity-weighted second moment of velocity by a factor of $2\sqrt{2\mathrm{ln 2}}$.
Figure \ref{fig:histo_sl_ps} (a), (b) and (c) show the histograms of the diameter, aspect ratio and velocity width in FWHM.
The minimum, maximum, mean value and standard deviation of each physical properties of identified cores are summarised in table \ref{tab:oriona_property}. The mean values of standard deviations of diameter, aspect ratio, and velocity width in FWHM of identified cores are $0.16\pm0.06$ pc, $0.58\pm0.16$, $0.33\pm0.14$ km s$^{-1}$, respectively.
Starless cores tend to have a small diameter and velocity width compare to protostellar cores. There are no clear differences between the aspect ratios of starless cores and protostellar cores.

\subsubsection{Core mass and virial mass}
\label{sec:mass_virial_ratio}

To estimate the core mass with minimizing the effect of the CO depletion, we used the \textit{Herschel}--\textit{Planck} H$_2$ column density data different from previous studies on the core identification based on C$^{18}$O observations \citep{ikeda07, ikeda09, shimajiri15}.
We determined the core mass by integrating the \textit{Herschel}--\textit{Planck} H$_2$ mass over the projected area of the core on the plane of the sky defined by Dendrogram on the C$^{18}$O image.
The dendrogram analysis allows us to distinguish between the cores and other structures. In other words, we can estimate the contribution of the emission associated with cores on the integrated intensity (see also figure 4 (left) and Sect 4.1.2 in \citep{rosolowsky08}).
When multiple pixels belonged to the different cores are overlapped along the line of sight at a pixel, we divide the \textit{Herschel}--\textit{Planck} H$_2$ mass between cores in proportion to the integrated intensity of C$^{18}$O emission after removing the contribution of the ambient gas.
Thus, the mass of the $k$-th core
is calculated as
\begin{equation}
        M_{\mathrm{core}}^k= \mu m_H \times \sum_i \left(N_{\mathrm{H_2},\,i}\times
    F_i^k \right)
 \label{eq:m_core}
\end{equation}
where
$F_i^k = {I_{i}^k}/{T_{i}}$ is the intensity fraction at the $i$-th pixel for the $k$-th core to sum of emission of all trunks,
$I_{i}^k$ is the C$^{18}$O integrated intensity at the $i$-th pixel for the $k$-th core, and $T_i (= \int_\mathrm{trunk} I_i dv)$ is the total velocity-integrated intensity at the $i$-th pixel in all trunk,
$\mu= 2.3$ is the mean molecular weight, $m_H$ is the mass of a hydrogen atom, $N_{\mathrm{H_2},\,i}$ is the \textit{Herschel}--\textit{Planck} H$_2$ mass contained in the $i$-th pixel.
We discuss the influence of the warm ambient gas (trunk) to this method in appendix \ref{app:leaf_trunk}.
About 60\% of identified cores overlaps with more than one core (see appendix \ref{app:overlap} for detail). Here we classified a pair of identified cores as overlapped cores when they overlaps more than one pixel along the line of sight.
The above expression can be rewritten as
\begin{equation}
    M_{\mathrm{core}}^k=4.15\times10^{-2}\,
            \left(\frac{\theta}{7.\arcsec5}\right)^2\, \left(\frac{D}{414\,\mathrm{pc}}\right)^2\,
            \left(\frac{\sum_i N_{\mathrm{H_2},\,i,\,k} \times  F_i^k}{10^{22}\,\mathrm{cm}^{-2}}\right)\,M_\odot
\end{equation}
The virial mass of a core is estimated as
\begin{equation}
    M_{\mathrm{vir}}=210\left(\frac{R_{\mathrm{core}}}{\mathrm{pc}}\right)
    \left(\frac{dV_{\mathrm{core}}}{\mathrm{km\,s^{-1}}}\right)^2M_\odot
\end{equation}
where $dV_{\rm core}$ is the FWHM velocity width and the core is assumed to be an uniform sphere and magnetic fields and external pressures are neglected for dynamical support.
The virial ratio is defined as
\begin{equation}
    \alpha_{\mathrm{vir}}=\frac{M_{\mathrm{vir}}}{M_{\mathrm{core}}}  \ .
\end{equation}
We note that for a centrally-condensed sphere with $\rho \propto r^{-2}$, the virial ratio becomes smaller by a factor of 5/3 than the above definition.
The cloud shape also influences the actual value of the virial parameter \citep{bertoldi92}.
If the CO depletion is significant particularly in cold dense regions, the estimates of the line-widths in the cold starless cores may be somewhat affected.

\begin{figure}[htbp]
 \begin{center}
  \includegraphics[scale=0.3,bb=0 0 1380 1380]{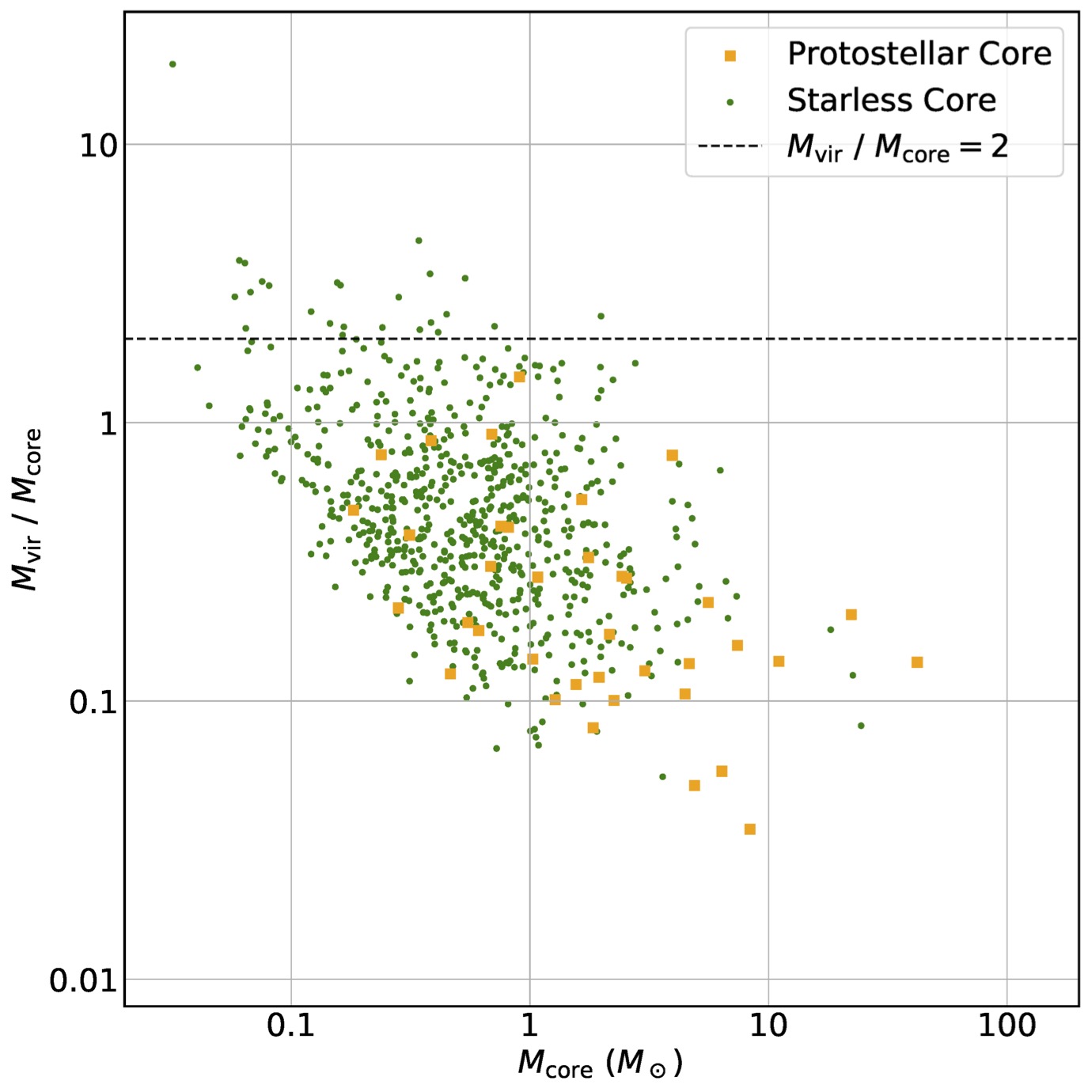}
 \end{center}
\caption{Virial ratio ($M_\mathrm{core}/M_\mathrm{vir}$) -- core mass relation. The horizontal dashed line shows $M_\mathrm{core}/M_\mathrm{vir}$=2. The starless cores and protostellar cores are shown as green dots and orange squares, respectively.}
\label{fig:virial_ratio}
\end{figure}

In this paper, we define the core whose virial ratio is smaller than 2 as a gravitationally bounded core.
Then, we identified 684 bound starless cores and 25 unbound starless cores, respectively.
All protostellar cores are likely to be bounded by gravity.
Figure \ref{fig:virial_ratio}, the virial ratio--mass relation, shows the clear trend that more massive cores tend to have smaller virial ratios.
The minimum, maximum, mean value, and the standard deviation of each physical properties of identified cores derived in this section are also summarised in table \ref{tab:oriona_property}.
The mean values of standard deviations of the mass, number density and virial ratio of identified cores are $1.08\pm 2.42$ M$_\odot$, $(0.87\pm1.92)\times 10^{4}$ cm$^{-3}$ and $0.64 \pm0.89$, respectively.

We assumed the local thermodynamic equilibrium (LTE) condition and optically thin emission of the C$^{18}$O ($J$=1--0)
to derive the mean column densities of the C$^{18}$O of individual cores.
Here, we assigned individual excitation temperature derived from $^{12}$CO observation (see \cite{nakamura19}) to each core.
The mean excitation temperature of projected areas of cores is 33 K.
Then we calculate the fractional abundance
of C$^{18}$O, $X_\mathrm{C^{18}O}$, for each core as a ratio of the column density of the C$^{18}$O and H$_2$.
The mean value is $\overline{X_\mathrm{C^{18}O}}$=5.7$\times$10$^{-7}$, which is 3 times larger than the representative value of $1.7\times 10^{-7}$ \citep{frerking82}.
Also, the number density of a core is obtained by assuming that the core has a constant density with a radius $R_{\mathrm{core}}$.
The physical properties of the identified starless and protostellar cores
such as core mass, virial ratio, density, and C$^{18}$O fractional abundance
are summarized in tables \ref{tab:catalog_position} and \ref{tab:catalog_property}, respectively.

Figures \ref{fig:histo_sl_ps} (d), (e) and (f) show the histograms of mass, virial ratio, and number density of the C$^{18}$O cores.
It is clear that protostellar cores are more massive and denser than starless cores.
A large fraction of identified cores have a virial ratio of smaller than 2 and protostellar cores tend to have smaller virial ratios.

\section{Discussion}
\label{sec:discussion}

\subsection{Comparison between the C$^{18}$O emission and dust emission}
\label{sec:c18o_h2}

\begin{figure}[htbp]
 \begin{center}
  \includegraphics[scale=0.15,bb=0 0 3140 1380]{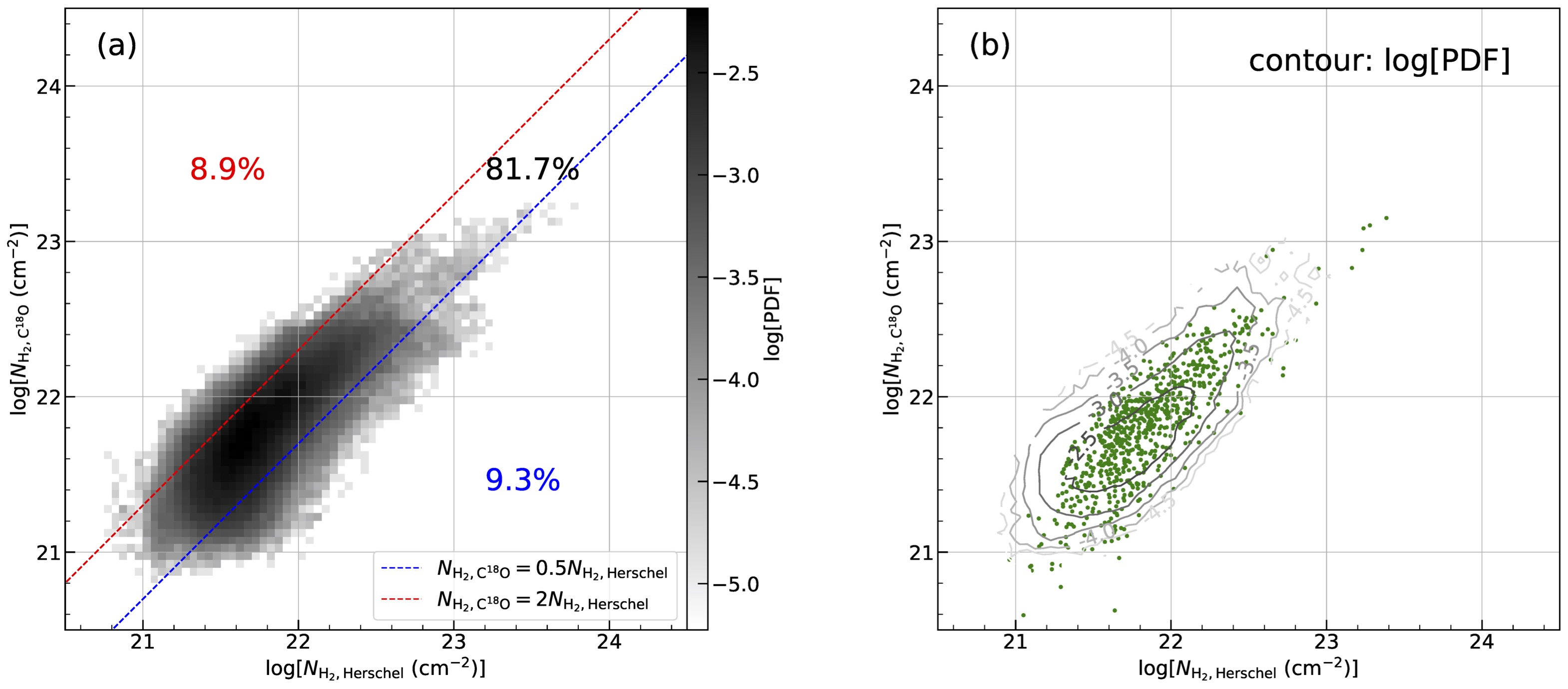}
 \end{center}
\caption{
Relationship between the H$_2$ column density derived directly from the C$^{18}$O integrated intensity, $N_\mathrm{H_2,C^{18}O}$, and the \textit{Herschel}--\textit{Planck} H$_2$ column density, $N_\mathrm{H_2,Herschel}$ is shown as (a) two dimensional histogram of PDF and (b) contour. Here we calculated values of pixels whose integrated intensity is larger than 15 local rms noise in the projections of all identified trunks onto the sky plane. The blue and red dashed lines in (a) show 2$N_\mathrm{H_2,C^{18}O}=N_\mathrm{H_2,Herschel}$ and $N_\mathrm{H_2,C^{18}O}=$2$N_\mathrm{H_2,Herschel}$. $N_\mathrm{H_2,C^{18}O}$ is the column density of H$_2$ calculated form the C$^{18}$O ($J$=1--0) observations and mean abundance ratio of C$^{18}$O. $N_\mathrm{H_2,Herschel}$ is the column density of H$_2$ derived from \textit{Herschel} observations. Each percentage represents the fraction of $N$ in each range: $N_\mathrm{H_2,C^{18}O}/N_\mathrm{H_2,Herschel}>$2 (red), 1/2$\leq N_\mathrm{H_2,C^{18}O}/N_\mathrm{H_2,Herschel}\leq$ 2 (black) and $N_\mathrm{H_2,C^{18}O}/N_\mathrm{H_2,Herschel}<$1/2 (blue). In (b), the contour is drawn from log[PDF]=-4.5 to -2.5 every log[PDF]=0.5.
The dots show all identified C$^{18}$O cores. The mean values of $N_\mathrm{H_2,C^{18}O}$ and $N_\mathrm{H_2,Herschel}$ in the projections of cores onto the plane of the sky are used to plot cores onto the figure.}
\label{fig:c18o-h2}
\end{figure}

Figure \ref{fig:c18o-h2} shows the relationship between the H$_2$ column density derived directly from the C$^{18}$O integrated intensity, $N_\mathrm{H_2,C^{18}O}$, and the \textit{Herschel}--\textit{Planck} H$_2$ column density, $N_\mathrm{H_2,Herschel}$, in Orion A.
The relationship is shown as a two-dimensional histogram of PDF (probability distribution function) in figure (a) and contour in figure (b), respectively.
Here, we obtained the C$^{18}$O column density under the assumption of LTE and optically-thin condition as \cite{ikeda09} and \cite{shimajiri15}.
The excitation temperature of C$^{18}$O was assumed to be the same as the peak intensity of $^{12}$CO emission.
We have calculated the C$^{18}$O column density of pixels which integrated intensity is larger than 15 local rms noise included in the projection of all trunks to the plane of the sky.
Then we converted the C$^{18}$O column density to the H$_{2}$ column density with a constant fractional abundance of C$^{18}$O to H$_{2}$ of $X_\mathrm{C^{18}O}$=5.7$\times$10$^{-7}$.
The plot indicates that the column density of C$^{18}$O is roughly proportional to the \textit{Herschel}--\textit{Planck} H$_2$ column density below $N_{H_2} \lesssim 5\times 10^{22}$ cm$^{-2}$.
In figure \ref{fig:c18o-h2} (a), the two dashed lines show $N_\mathrm{H_2,C^{18}O}=0.5N_\mathrm{H_2,Herschel}$ and $N_\mathrm{H_2,C^{18}O}=2N_\mathrm{H_2,Herschel}$.
For larger H$_2$ column density ($n \gtrsim 10^5$ cm$^{-3}$),
the C$^{18}$O column density tends to level off for larger column density.
The number of such pixels is only 18.2 \%.
In figure \ref{fig:c18o-h2} (b), the identified cores are plotted onto the contour.
For most of the cores identified, the C$^{18}$O column density is more or less proportional to the \textit{Herschel}--\textit{Planck} H$_2$ column density.
Thus, the C$^{18}$O is likely to be a reliable tracer of the molecular hydrogen mass with the exception of cold and dense regions with $N_{H_2} \gtrsim 5\times 10^{22}$ cm$^{-2}$.

For dense regions, C$^{18}$O appears to be less abundant.
Therefore, the cores located in the dense regions may not be well traced by C$^{18}$O emission.
However, the number fraction of such cores is likely to be small for the entire Orion A. There may be at least two reasons why C$^{18}$O abundance is lower toward denser regions. One is the CO depletion on to the grain surfaces for cold ($T\lesssim 20 $ K), dense ($n \gtrsim 10^5$ cm$^{-3}$) parts. Another is the dissociation due to FUV radiation \citep{shimajiri14,lin16,ishii19}. In Orion A, several massive stars, mainly located in the OMC-1 region, emit strong UV radiation, which can dissociate C$^{18}$O in dense regions where $^{12}$CO and $^{13}$CO emissions become optically thick.

Our result is consistent with that of \citet{ripple13} with $^{13}$CO. They showed that
the effect of CO freeze out is limited to regions with $N_{H_2} \gtrsim 10^{22}$ cm$^{-2}$ (Av $\gtrsim$ 10 mag) and gas temperatures less than
$\sim 20$ K for Orion.

\subsection{Consideration of the effect of the ambient gas to mass estimation}
\label{sec:ambient_gas}

\begin{figure}[htbp]
 \begin{center}
  \includegraphics[scale=0.25,bb=0 0 1744 1656]{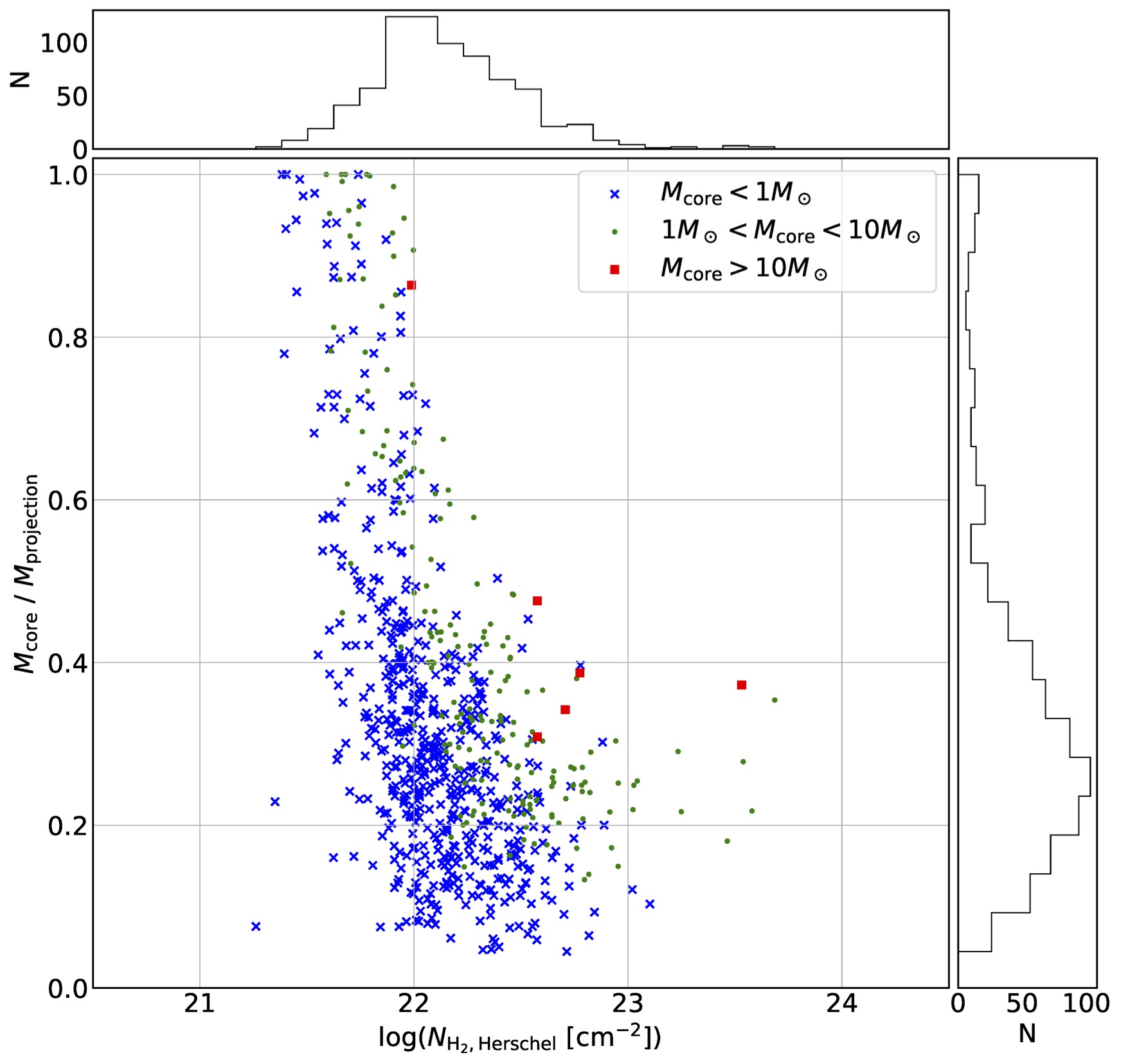}
 \end{center}
\caption{
Mass ratio $M_\mathrm{core}$ \ $M_\mathrm{projection}$ -- $N_\mathrm{H_2,Herschel}$ relation. To calculate $M_{projection}$, all H$_2$ column density is assigned to cores. When multiple cores overlap along the line of sight, H$_2$ column density proportional to each intensity is distributed to each core.
The mean value and standard deviation of mass ratio $M_\mathrm{core}$ \ $M_\mathrm{projection}$ are 0.35$\pm$0.21.
The blue crosses, green circles and red squares show the cores whose masses are $M_\mathrm{core}<1M_\odot$, $1M_\odot<M_\mathrm{core}<10M_\odot$ and $M_\mathrm{core}>10M_\odot$, respectively.}
\label{fig:massratio}
\end{figure}

In this section, we compare the core masses calculated with two different approaches.
One is the core mass weighted by the C$^{18}$O emission ($M_\mathrm{core}$) defined by equation (\ref{eq:m_core}) in section \ref{sec:mass_virial_ratio}.
The other ($M_\mathrm{projection}$) is derived by summing up all \textit{Herschel}--\textit{Planck} H$_2$ column density included in the projected area of a core to the plane of the sky without removing the ambient components of the gas.
This is essentially the same as the mass estimation based on some core identification schemes such as \texttt{clumpfind} \citep{williams94} and Fellwaker \citep{berry15}.
When multiple cores overlap along the line-of-sight, we distribute the \textit{Herschel}--\textit{Planck} H$_2$ column density to each core.
The assigned H$_2$ column density of each core is proportional to the C$^{18}$O intensity of a core.
The difference between $M_\mathrm{core}$ and $M_\mathrm{projection}$ is the treatment of ambient gas of the dense core.
$M_\mathrm{projection}$ is a sum of $M_\mathrm{core}$ and mass of the ambient gas along the line of sight.
Then we calculated mass ratio, $M_\mathrm{core}/M_\mathrm{projection}$ and show the mass ratio -- \textit{Herschel}--\textit{Planck} H$_2$ column density relation in figure \ref{fig:massratio}.
As shown in this figure, an identified core with a higher H$_2$ column density tends to have a smaller mass ratio.
This suggests that such cores have more ambient gas and the core mass derived from dust observations tends to be more overestimated.
A large fraction of core has mass ratios of smaller than 0.5 and the mean mass ratio and standard deviation are 0.35$\pm$0.21.
Then, more than 50\% of H$_2$ column density seems to come from the ambient gas in the dense regions whose H$_2$ column density is $\gtrsim 10^{22}$ cm$^{-2}$.
In other words, the mass estimated only from the 2D dust emission map is about 3 times larger than the actual mass evaluated in 3D.
This implies that the core masses estimated with the dust emission and extinction are significantly larger than the actual values.
For further discussion, we need more observations with other molecular lines.

\subsection{Core mass function in Orion A}
\label{sec:cmf}

In this section, we derive the CMF in Orion A using the cores identified above and $M_\mathrm{core}$ and then discuss their properties.
Here, the CMF is defined as the number of cores with masses in a mass range from $M_\mathrm{core}$ to
$M_\mathrm{core}+ d M_\mathrm{core}$.

\begin{figure}[htbp]
 \begin{center}
  \includegraphics[scale=0.3,bb=0 0 1380 1380]{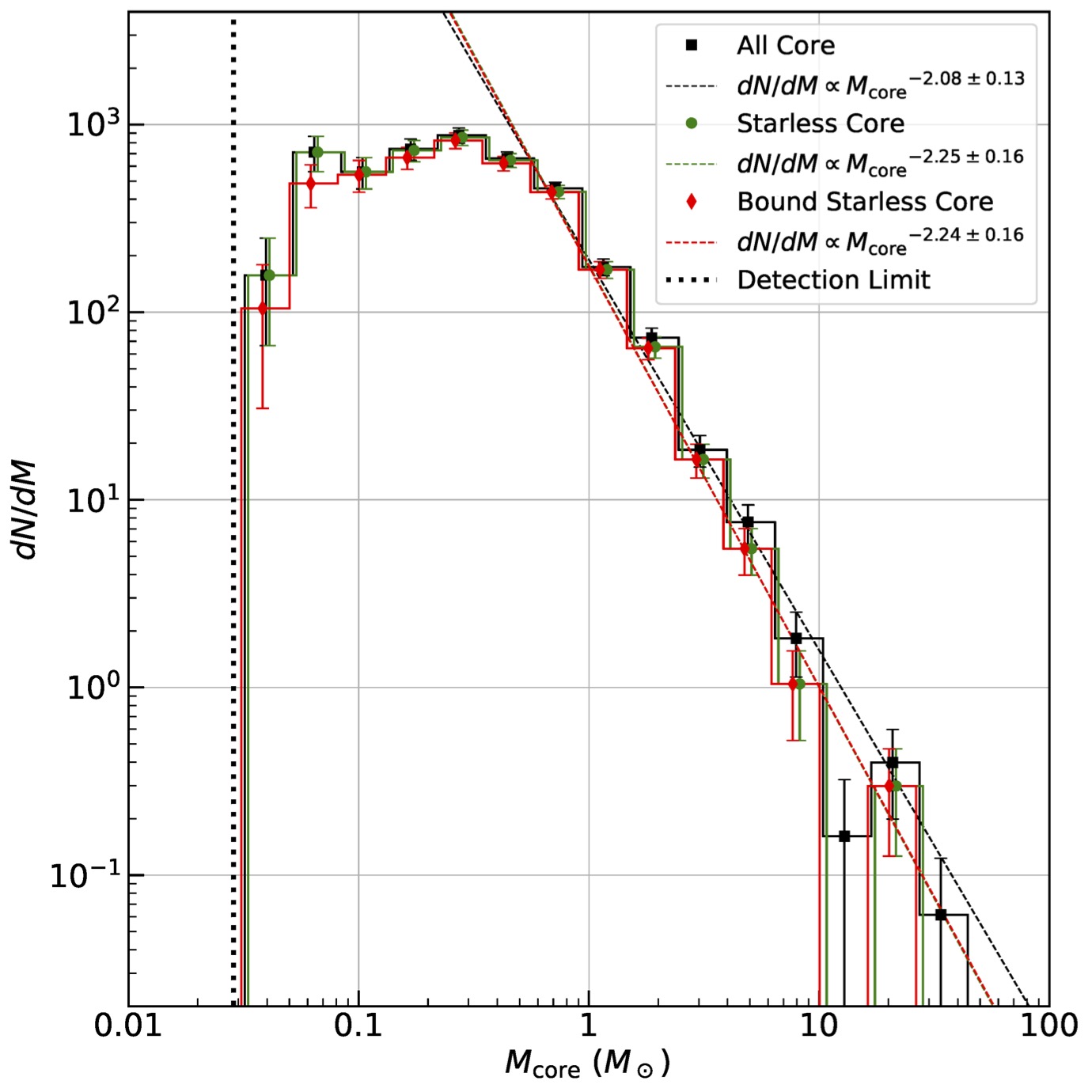}
 \end{center}
\caption{CMFs for the identified cores (black line), starless cores (green line), and gravitationally bound starless cores (red line). The error bars show the statistical uncertainty calculated as the square root of the number of cores in each mass bin, $\sqrt N$. The dashed line shows the best-fit single power-law functions for each CMF between one mass bins above the turnover and the high-mass end. The dotted line shows the detection mass limit of $\sim 0.03M_\odot$.}
\label{fig:cmf}
\end{figure}

Figure \ref{fig:cmf} shows the CMFs for all identified cores, starless cores, and gravitationally bound cores.
Since almost all identified cores seem to be bounded by gravity as shown in Section \ref{sec:mass_virial_ratio}, we do not classify bound and unbound cores to derive CMFs in this paper.
The mass detection limit for our analysis is $\sim 0.03M_\odot$.
This mass is derived from the minimum intensity and size to pass the core identification conditions
described in section \ref{sec:core_id} and assumptions of LTE condition and optically thin emission of C$^{18}$O ($J$=1--0) emission.
Here we applied $T_\mathrm{ex}$=33 K and $X_\mathrm{C^{18}O}$=5.7$\times$10$^{-7}$.
Both CMFs have turnovers at $\sim$0.3 $M_\odot$ and power-law like shape above them.
The best-fit power-law indexes of CMFs for all cores, starless cores, and gravitationally bound cores are $-$2.08, $-$2.25, and $-$2.24, respectively.
The best-fit power-law indices are in agreement with that of the Salpeter IMF ($-$2.35, \cite{salpeter55}).
The slopes in the CMFs are also consistent with the previous study \citep{ikeda09}
that identified C$^{18}$O cores with a different core identification scheme \texttt{clumpfind} in a much smaller area including the OMC-1 region ($\sim$2.4 pc $\times$ 2.4 pc).
Their C$^{18}$O data have the same effective angular resolution and velocity resolution as ours.
However, a turnover mass of their CMF for the OMC-1 region, $\sim 5 M_\odot$,
is about three times larger than our value.
This difference mainly comes from the usage of the different core identification schemes. For the \citet{ikeda07}'s scheme, i.e., \texttt{clumpfind}, the core mass tends to be larger than that obtained based on Dendrogram since for the former, all the pixels located inside the closed contour which contains cores are assigned to the adjacent core.

In order to estimate the completeness of core identification, we derived the detection probability as a function of the core mass by putting the artificial cores generated as follows.
\begin{enumerate}
  \item We decided the mass of each artificial core as the central value of each mass bin on the log scale.
  \item With the assumptions of the LTE condition and optically thin emission of the C$^{18}$O ($J$=1--0), we converted the mass of each artificial core to the integrated intensity of C$^{18}$O emission using $T_\mathrm{ex}$=33 K and $X_\mathrm{C^{18}O}$=5.7$\times$10$^{-7}$.
  \item We calculated the radius and velocity width of each core using the radius -- mass and velocity width in rms -- mass relations, which are derived by fitting for all identified cores.
  Here we derived radius as the geometric mean of major and minor axes of identified cores. The mean values of major and minor axes are shown in table \ref{tab:catalog_position}.
  \item We made a modified dataset by inserting the artificial cores that have the three-dimensional Gaussian shape in the p-p-v space with sizes and velocity widths derived above. Here we just added the artificial
  cores to the observed data. The positions and the system velocities of the artificial cores that correspond with individual mass bins are given randomly in the
  trunks which are the lowest structures of the Dendrogram's hierarchies.
  \item We applied the same core identification method to the modified data and checked
  whether the artificial core is identified or not.
  \item We repeated this procedure $N_\mathrm{trial}$ times for each each mass bin in the CMF plot and calculated the possibility
  that the artificial cores are identified as independent cores $F(M_\mathrm{LTE}/M_\odot)$ with
  equation (\ref{eq:id_probability}).
\end{enumerate}
To calculate the probability that the artificial cores are identified as independent
cores, first, we prepared two data; the original observed data and modified data which contains
the artificial core. We applied the same core identification method described in Section
\ref{sec:core_id}. Then, we counted the number of identified cores within 1$\sigma$ from the center
of the artificial core in observed data $n_\mathrm{observed}$ and modified data
$n_\mathrm{artificial}$. We summed up the number of the case of
$n_\mathrm{artificial}-n_\mathrm{observed}\geq 1$ for $N_\mathrm{trial}$ = 100 trials as $n_\mathrm{identified}$.
Finally, the probability function that the artificial cores are identified as independent cores is
derived as
\begin{equation}
    P(M_\mathrm{LTE}/M_\odot)=\frac{n_\mathrm{identified}(M_\mathrm{LTE}/M_\odot)}{N_\mathrm{trial}}
    \label{eq:id_probability}
\end{equation}
Then, the corrected number of cores in each mass bin $N_\mathrm{corrected}$ is calculated
with the observed number of cores $N_\mathrm{observed}$ as
\begin{equation}
    N_\mathrm{corrected}=\frac{N_\mathrm{observed}}{P(M_\mathrm{LTE}/M_\odot)}
    \label{eq:n_compensated}
\end{equation}

\begin{figure}[htbp]
 \begin{center}
  \includegraphics[scale=0.3,bb=0 0 1380 1380]{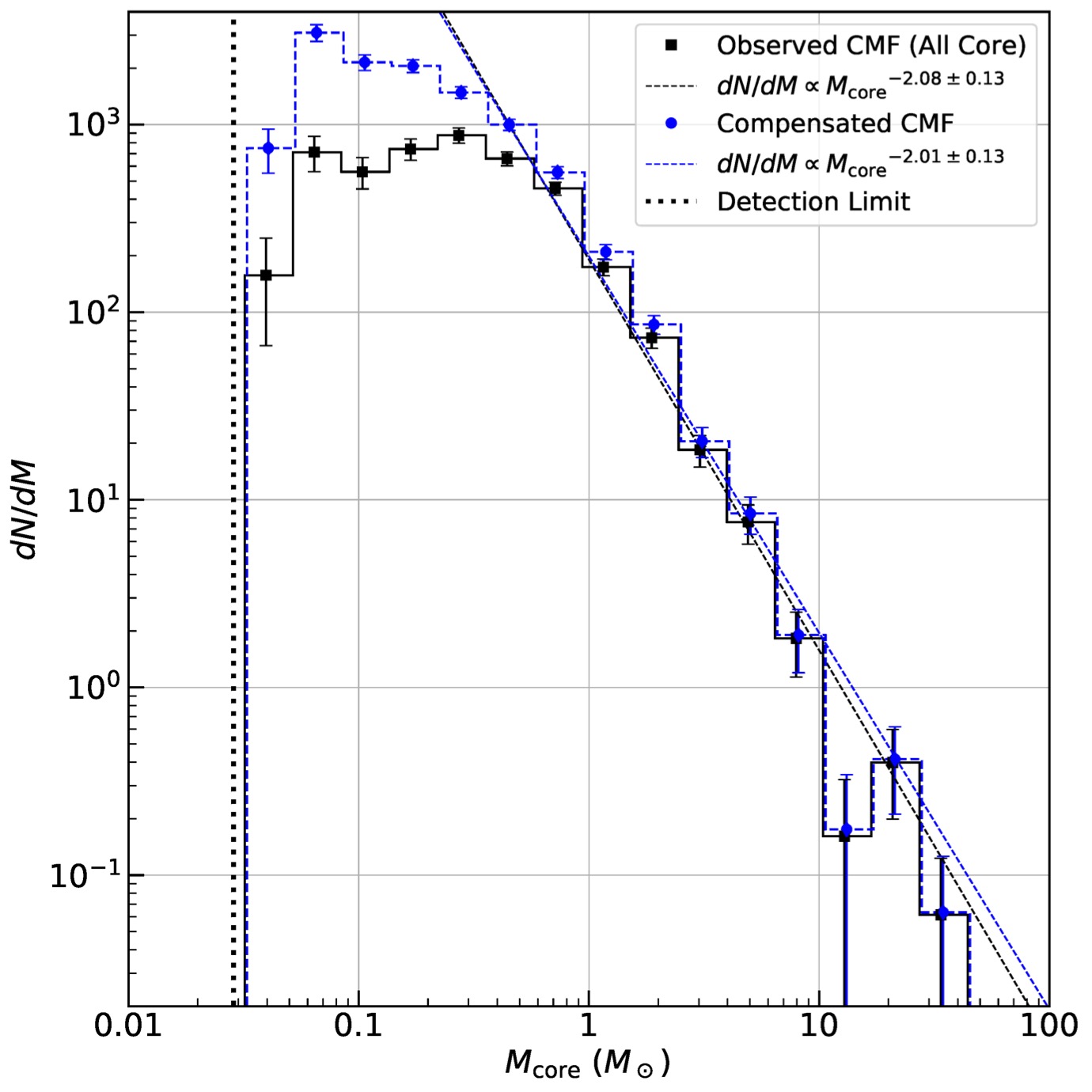}
 \end{center}
\caption{Observed CMF (black line) and compensated CMF (blue line) for the identified cores. In compensated CMF, we derived the number of cores in each mass bin by dividing the observed number by detection probability.}
\label{fig:cmf_comp}
\end{figure}

Figure \ref{fig:cmf_comp} shows the observed and compensated CMFs
The shape of the compensated CMF at the high-mass end is essentially the same as the observed one.
In contrast, we see significant differences between the distribution at low-mass parts of observed CMF and compensated CMF.
The number of low-mass cores in observed CMF is smaller than that in compensated CMF and such cores are likely to be missed in observed CMF.
We find no clear turnover in the compensated CMF.
Thus, it is difficult to conclude that
the CMFs have a turnover at the low-mass part from our observations.
This is consistent with the argument by \citet{ikeda07}
This fact suggests that the spatial resolution of our data is not enough to confirm whether a CMF has a turnover and we need a higher spatial resolution to address this point.

\section{Comparison with the CMF derived from the dust emission}
\label{sec:dust_c18o}

\begin{figure}[htbp]
 \begin{center}
  \includegraphics[scale=0.3,bb=0 0 1376 1378]{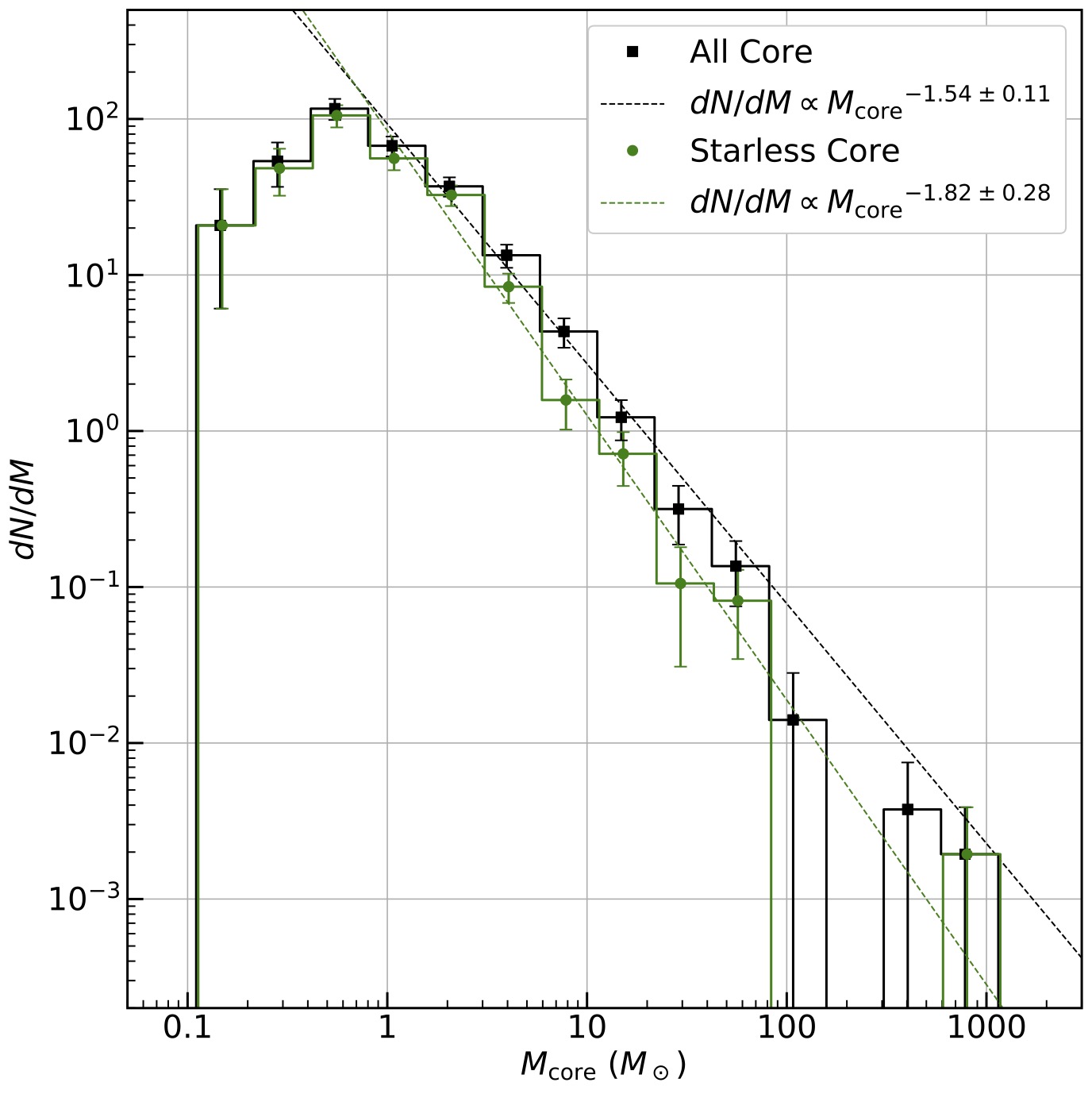}
 \end{center}
\caption{CMFs of all dust cores and starless cores identified in (nutter07) in Orion A. The dashed line shows the best-fit single power-law functions for each CMF above the turnover as figure \ref{fig:cmf}.}
\label{fig:cmf_nutter07}
\end{figure}

In section \ref{sec:ambient_gas}, we pointed out that the masses derived from the dust emission map tend to be significantly overestimated due to the contribution of the ambient gas. Here, we attempt to assess how this overestimation affects previous results by using \citet{nutter07}'s core catalog.
Figure \ref{fig:cmf_nutter07} shows the CMF derived from the SCUBA 850 $\mu$m emission map with a angular resolution of $\sim 15$\arcsec.
The core identification scheme of \citet{nutter07} is also different from ours. They used not a hierarchy but a signal-to-noise ratio and a detailed explanation are in the paper.
The authors set the distance of Orion A as 400 pc and we recalculated the core masses with a distance of 414 pc.
As a result, the turnover mass derived from the dust emission ($\sim$ 0.5 M$_\odot$) appears to be comparable to our value obtained from the C$^{18}$O emission.
However, the dust mass tends to be overestimated the core mass since the dust mass sums up both the core mass and the ambient gas overlapped along the line of sight. This is one of the disadvantages to estimate the core masses only from the dust observations.
It is worth noting that in the SCUBA dust image,
the structures larger than $\sim$ 1.'5 are removed during the data reduction process. This artificial effect leads to underestimates of the dust emission.

\begin{figure}[htbp]
 \begin{center}
  \includegraphics[scale=0.18,bb=0 0 2500 2497]{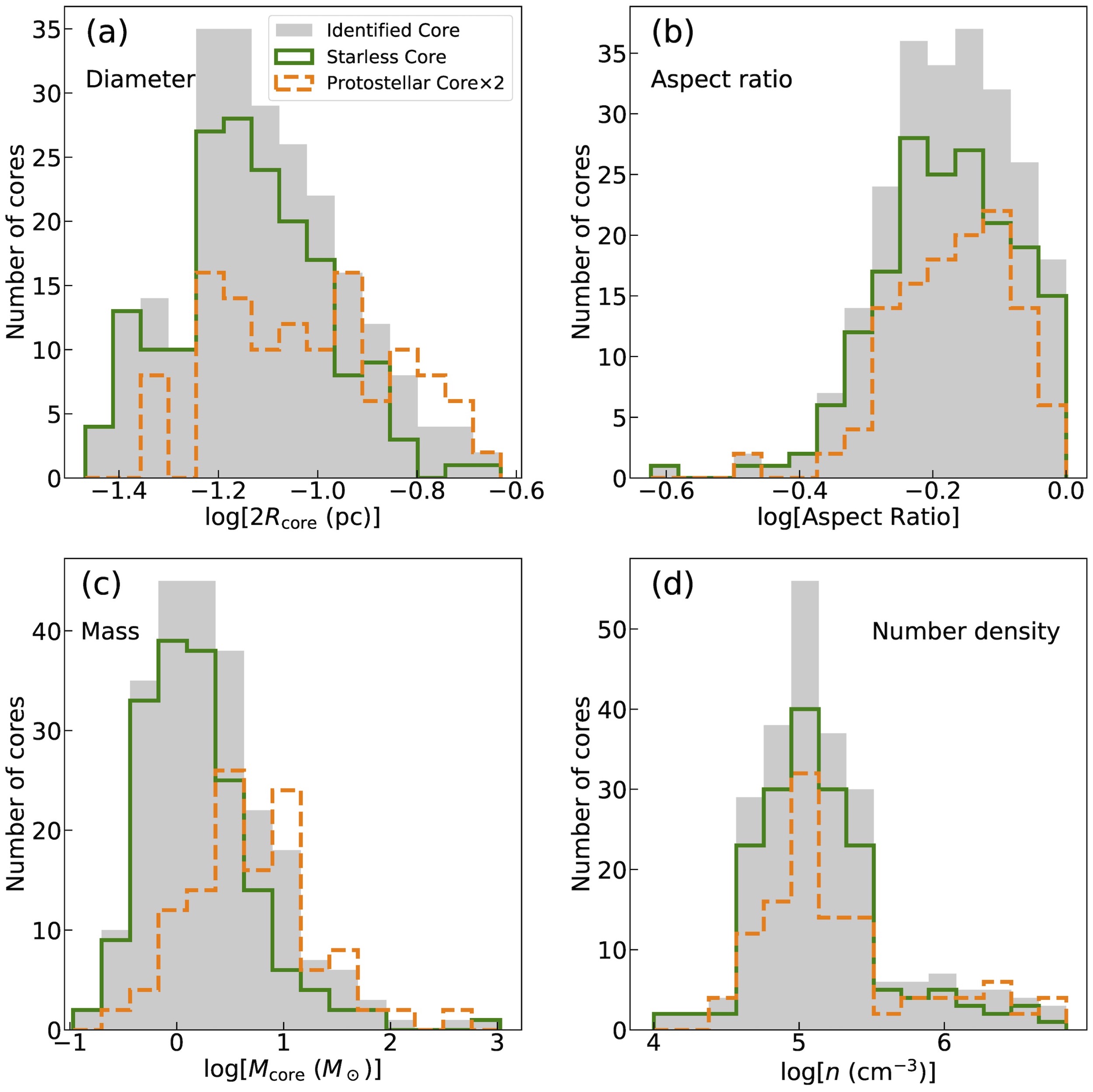}
 \end{center}
\caption{The histograms of (a) the diameter (b) the aspect ratio, (c) the core mass, (d) the number density of dust cores, respectively.
The histogram for the all dust cores is shown in grey.
The green and orange histograms are for starless cores and protostellar cores.}
\label{fig:histo_nutter07}
\end{figure}

In addition, the cores overlapped along the line of sight are difficult to be distinguished only from the dust map.
In fact, the most massive dust core in \citet{nutter07} has a mass of $\sim$ 1062 M$_\odot$, located in OMC-1.
In our C$^{18}$O case, the most massive core has a mass of $\sim$ 42 M$_\odot$.
In the area containing the most massive dust core, our C$^{18}$O data indicate the existence of the multiple cores overlapped and/or a lot of ambient gas surrounding cores.
This effect makes massive cores more massive and the slope of the CMF shallower toward the high-mass part, and this is another disadvantage for the dust observations.
Figure \ref{fig:histo_nutter07} shows the histograms of physical properties of dust cores such as the core diameter, aspect ratio, mass, and density.
Although the mass ranges of C$^{18}$O cores and dust cores are similar, dust cores tend to be smaller than C$^{18}$O cores.
Then, dust cores are likely to be denser than C$^{18}$O cores.
There are no big differences in aspect ratio between the two cores.

As derived in section \ref{sec:ambient_gas}, the dust mass may be overestimated by a factor of 3 from our data.
If we apply this correction factor to their CMF, the turnover mass becomes small at 0.1--0.2 M$_\odot$.
This turnover mass is comparable to that of the stellar IMF in Orion Nebular Cluster \citep{hillenbrand97, dario12}.
If we apply the same discussion of \citet{alves07},
this may indicate that star formation efficiency (SFE) of individual cores is much higher than the prediction of the theoretical studies \citep{matzner00, machida12}, in which SFE is estimated to be 30--50\% since the protostellar outflows blow out of a significant amount of core mass.
Such a high SFE appears to be consistent with the result by \citet{bontemps01}.
We note that recent numerical simulations have suggested that the mass infall from the surrounding play an important role in the evolution of the core mass \citep{vazquez19,padoan20}. Such processes also affect the peak masses of CMFs i.e., the turnover mass would evolve over time due to the mass accretion.

Considering the overestimation, the turnover mass is expected to be $\sim$ 0.1 $M_\odot$, and observations with a higher angular resolution are needed to confirm it.
In a forthcoming paper, we attempt to identify C$^{18}$O cores in Orion A using the CARMA-NRO combined data \citep{kong18} with a much higher angular resolution ($\sim 8"$) to reveal the shape of the CMF at low mass part.



\section{Summary}
\label{sec:summary}
We conducted an unbiased dense core survey in Orion A with the C$^{18}$O data taken by Nobeyama 45-m telescope.
The main conclusions are summarized as follows.
\begin{itemize}
\item[1.] The C$^{18}$O($J$=1--0) emission is likely to be a reliable tracer of molecular mass toward the regions with $N \lesssim 5\times 10^{22}$ cm$^{-2}$. The average abundance of C$^{18}$O relative to H$_2$ is obtained to be $5.7\times 10^{-7}$. Its abundance tends to be smaller for the denser region within a factor of a few.
\item[2.] Applying Dendrogram, we conducted hierarchical structure analysis of the Orion A cloud and define leaves as cores. As a result, we identified 746 cores.
With the HOPS catalog, we classified the cores into two groups:
709 starless cores and 37 protostellar cores.
\item[3.] We derived the masses of the cores identified from the \textit{Herschel}--\textit{Planck} H$_2$ map, by removing the contribution of the ambient gas that is not associated with cores along the line of sight.
\item[4.] The mean diameter, mass, and velocity widths of the starless cores are estimated to be 0.16 $\pm$ 0.05 pc,
0.92 $\pm$ 1.65 $M_\odot$, and 0.32 $\pm$ 0.14 km s$^{-1}$, respectively.
\item[4.] The CMF of the starless bound cores has a slope similar to the Salpeter IMF in the high-mass part
(a power index of $\alpha = -$2.24 $\pm$ 0.16 at $\gtrsim$ 0.44 $M_\odot$).
and has a peak at 0.3 M$_\odot$.
\item[5.] The test of completeness indicates that this peak mass may be affected by the angular resolution of the map, and the actual peak mass, if exists, would be smaller than the observed value from our data.
\item[6.] The estimated mass of cores based on the dust observations is likely to be overestimated the actual core mass, presumably be a factor of $\sim 3$.
\item[7.] If we adopt this correction factor for the CMF obtained by \citet{nutter07}, the peak mass in the CMF tends to be comparable to the stellar IMF.
\end{itemize}


\begin{ack}
This work was carried out as one of the large projects of the Nobeyama
Radio Observatory (NRO), which is a branch of the National Astronomical
Observatory of Japan, National Institute of Natural Sciences.
We thank the NRO staff for both operating the 45 m and helping us with the data reduction.
Data analysis was carried out on the Multi-wavelength Data Analysis System operated by the Astronomy Data Center (ADC),
National Astronomical Observatory of Japan.
P.S. was partially supported by a Grant-in-Aid for Scientific Research (KAKENHI Number 18H01259) of Japan Society for the Promotion of Science (JSPS).
This work was supported in part by The Graduate University for Advanced Studies, SOKENDAI.
We thank the anonymous referee for many useful comments that have improved the presentation.
\end{ack}

\bibliographystyle{pasj}
\bibliography{nro45_paper.bib,nakamura.bib}

\clearpage

\appendix

\section{H$_2$ column densities of leaf and trunk with different temperature}
\label{app:leaf_trunk}
\begin{figure}[htbp]
 \begin{center}
  \includegraphics[scale=0.2,bb=0 0 2048 1536]{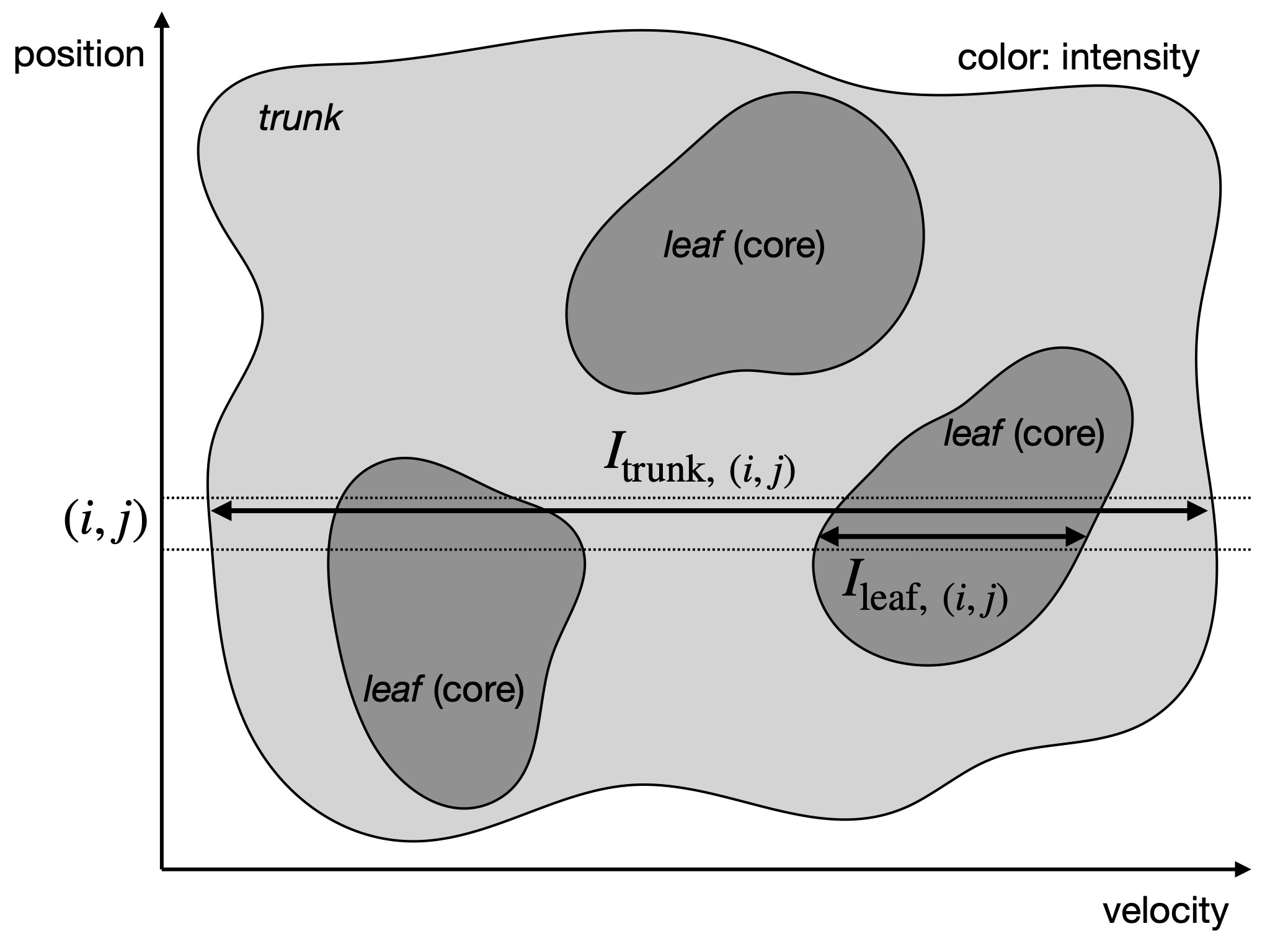}
 \end{center}
\caption{Schematic representation of how to calculate the core masses using the \textit{Herschel}--\textit{Planck} H$_2$ column density and C$^{18}$O intensity maps.}
\label{fig:cartoon_mass}
\end{figure}

\begin{figure}[htbp]
 \begin{center}
  \includegraphics[scale=0.15,bb=0 0 2818 1380]{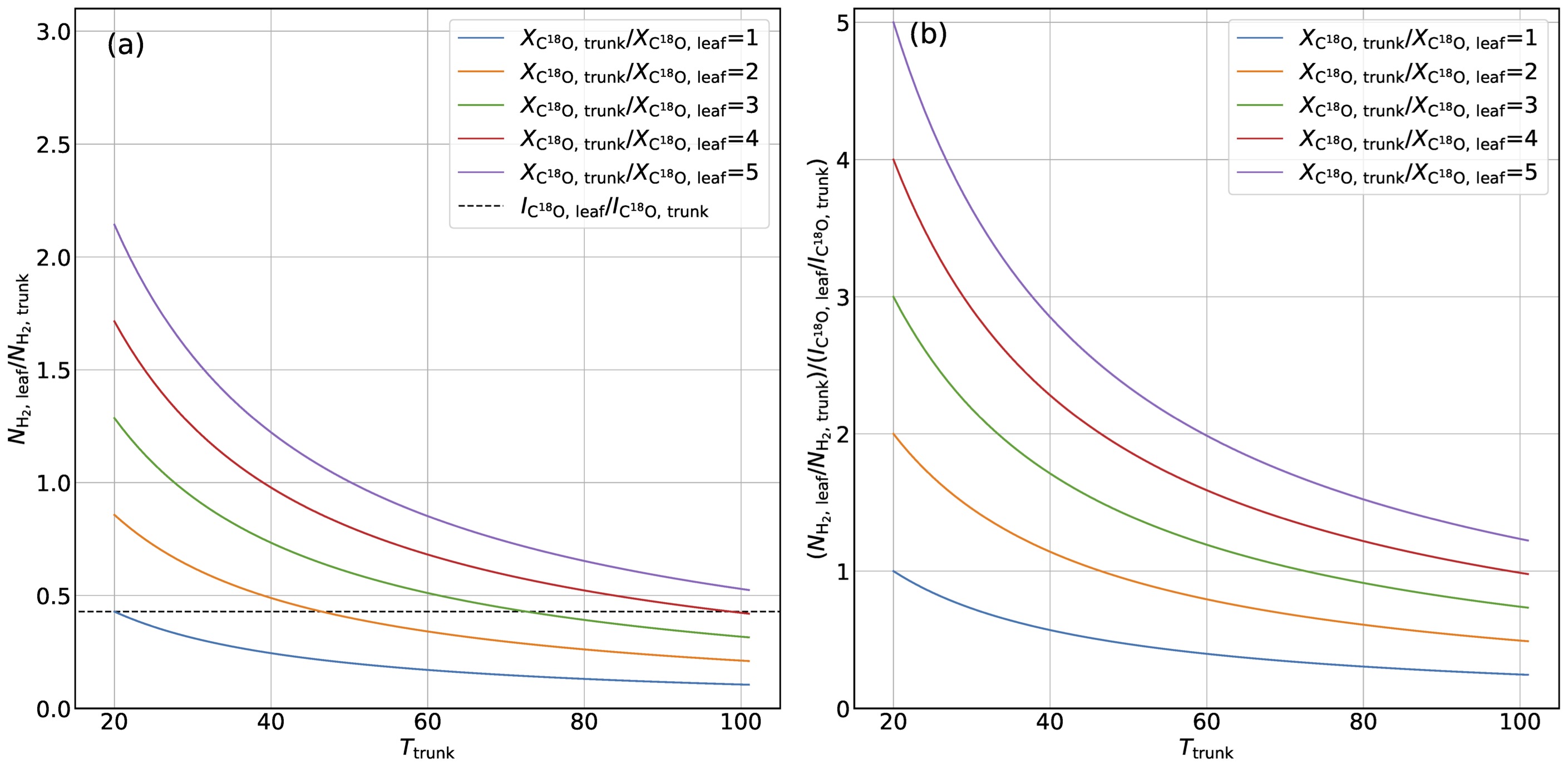}
 \end{center}
\caption{(a) The relationship between H$_2$ column density ratio, $N_\mathrm{H_2,\,leaf}/N_\mathrm{H_2,\,trunk}$ and temperature of trunk with the abundance ratio, $X_\mathrm{C^{18}O,\,trunk}/X_\mathrm{C^{18}O,\,leaf}$, of form 1 to 5. Here we fix the temperature of leaf and integrated intensity ratio, $I_\mathrm{C^{18}O,\,leaf}/I_\mathrm{C^{18}O,\,trunk}$, as 20 K and 3/7 (from section \ref{fig:massratio}), respectively.
(b) The relationship between ratio of H$_2$ column density ratio to integrated intensity ratio and temperature of trunk.}
\label{fig:leaf_trunk}
\end{figure}

In equation \ref{eq:m_core}, we derived core mass with C$^{18}$O intensity ratio, $I_\mathrm{C^{18}O,\,leaf}/I_\mathrm{C^{18}O,\,trunk}$ (Figure \ref{fig:cartoon_mass}), of leaves and trunks by assuming that leaves and trunks have the same temperature and fractional abundance of C$^{18}$O to H$_2$.
In the real situation, however, cores and ambient gas seem to have different temperature and fractional abundance i.e., ambient gas has a higher temperature and larger abundance ratio than cores.
In this section, we investigate the influence of such differences on core mass estimation.
To do so, we calculate the ratio of H$_2$ column density of leaf and trunk from the intensity ratio of leaves and trunks without the assumption that leaves and trunks have the same temperature and fractional abundance using a simple model. First, we fixed the intensity ratio of leaf and trunk and the temperature of leaf as 3:7 (see section \ref{sec:ambient_gas}) and 20 K. Second, we calculated C$^{18}$O column density ratio of leaf and trunk, $N_\mathrm{C^{18}O,\,leaf}/N_\mathrm{C^{18}O,\,trunk}$, from the temperature of trunk is 20 K to 100 K. Finally, we converted the C$^{18}$O column density ratio to H$_2$ column density ratio, $N_\mathrm{H_2,\,leaf}/N_\mathrm{H_2,\,trunk}$, with the ratio of fractional abundance ratio, $X_\mathrm{C^{18}O,\,trunk}/X_\mathrm{C^{18}O,\,leaf}$, from 1 to 5.
Figure \ref{fig:leaf_trunk} (a) shows the relationship between H$_2$ column density ratio and the temperature of trunk. In figure \ref{fig:leaf_trunk} (b), we normalized H$_2$ column density ratio with intensity ratio.
When trunks (ambient gas) have higher temperature compare to leaves (cores), H$_2$ column density ratio, $N_\mathrm{H_2,\,leaf}/N_\mathrm{H_2,\,trunk}$, is larger than intensity ratio and we underestimate the core masses.
In contrast, when the abundance ratio of a trunk is larger than that of a leaf, H$_2$ column density ratio becomes larger.
For example, when $T_\mathrm{trunk}$=40 K (=2$T_\mathrm{leaf}$) and $X_\mathrm{C^{18}O,\,trunk}/X_\mathrm{C^{18}O,\,leaf}$=2, H$_2$ column density ratio is similar to C$^{18}$O intensity ratio.
We expect both effects happen simultaneously in the real situation and we think we can derive H$_2$ column density ratio from C$^{18}$O intensity ratio of a leaf and a trunk to calculate core mass with equation \ref{eq:m_core}.

\section{Overlapping cores}
\label{app:overlap}
\begin{figure}[htbp]
 \begin{center}
  \includegraphics[scale=0.15,bb=0 0 2960 1380]{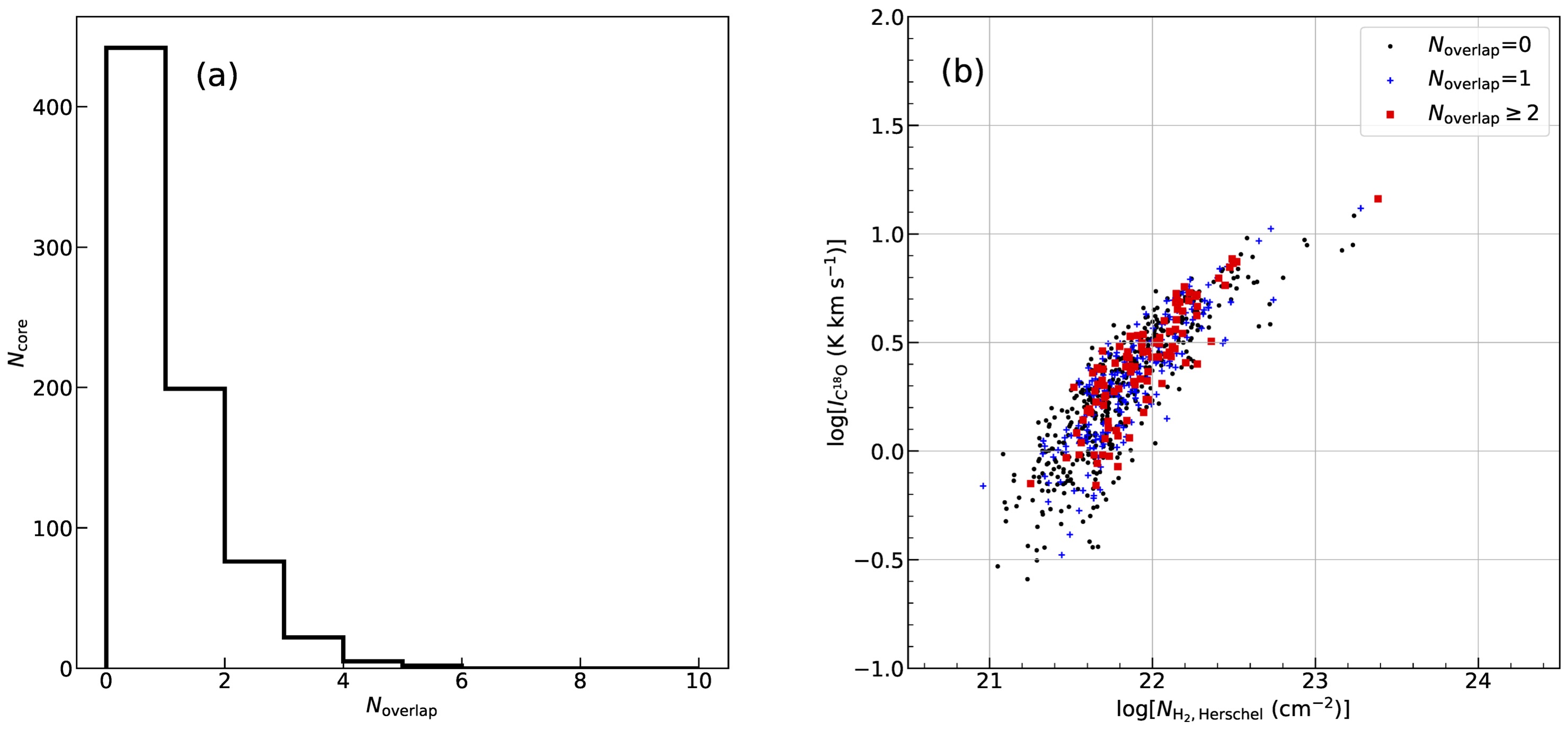}
 \end{center}
\caption{(a) The histogram of the number of overlapped cores. (b) The distribution of overlapped cores in $I_\mathrm{C^{18}O}$--$N_\mathrm{H_2,\,leaf}$ plane.}
\label{fig:n_overlap}
\end{figure}

As we mentioned in section \ref{sec:mass_virial_ratio}, around 60\% of identified cores overlap each other.
In this section, we summarize the properties of overlapped cores such as the number of them, integrated C$^{18}$O intensity, and H$_2$ column density.
Figure \ref{fig:n_overlap} (a) shows a histogram of the number of overlapped cores, $N_\mathrm{overlap}$,. For more than 60\% of overlapped cores, $N_\mathrm{overlap}$=1 and several cores have $N_\mathrm{overlap}$ of larger than 3.
Figure \ref{fig:n_overlap} (b) represents the distributions of overlapped cores in $I_\mathrm{C^{18}O}$--$N_\mathrm{H_2,\,leaf}$ plane and there is no bias of the distribution. Then, we do not recognize any critical integrated intensity or column density which overlapping effect becomes severe.

\clearpage

\clearpage

\begin{table}[htbp]
 \centering
  \caption{The Results of Core Identification in Orion A}
\begin{tabular}{cccc}
 \hline \hline 
 Category & N\\ 
 \hline 
 Identified Core & 746 \\ 
 Starless Core & 709 \\ 
 Protostellar Core & 37 \\ 
 \hline 
\end{tabular}
  \label{tab:oriona_core_id}
\end{table}

\begin{table}[htbp]
 \centering
  \caption{The Summary of Core Properties in Orion A}
  \small
  \begin{tabular}{ccccc}
 \hline \hline 
 Property & Cetegory & Minimum & Maximum & Mean $\pm$ Std. \\ 
 \hline 
  & Identified Core & 0.07 & 0.62 & 0.16 $\pm$ 0.06\\ 
Diameter (pc) & Starless Core & 0.07 & 0.62 & 0.16 $\pm$ 0.05\\ 
  & Protostellar Core & 0.08 & 0.50 & 0.20 $\pm$ 0.10\\ 
 \hline 
  & Identified Core & 0.20 & 0.98 & 0.58 $\pm$ 0.16\\ 
Aspect Ratio & Starless Core & 0.20 & 0.98 & 0.58 $\pm$ 0.16\\ 
  & Protostellar Core & 0.24 & 0.92 & 0.56 $\pm$ 0.16\\ 
 \hline 
  & Identified Core & 0.12 & 0.96 & 0.33 $\pm$ 0.14\\ 
FWHM (km s $^{-1}$) & Starless Core & 0.12 & 0.96 & 0.32 $\pm$ 0.14\\ 
  & Protostellar Core & 0.18 & 0.81 & 0.36 $\pm$ 0.16\\ 
 \hline 
  & Identified Core & 0.03 & 42.00 & 1.08 $\pm$ 2.42\\ 
Mass (M$_\odot$) & Starless Core & 0.03 & 24.45 & 0.92 $\pm$ 1.65\\ 
  & Protostellar Core & 0.18 & 42.00 & 4.12 $\pm$ 7.51\\ 
 \hline 
  & Identified Core & 0.02 & 37.36 & 0.87 $\pm$ 1.92\\ 
Number Density (10$^{4}$ cm$^{-3}$) & Starless Core & 0.02 & 37.36 & 0.84 $\pm$ 1.94\\ 
  & Protostellar Core & 0.12 & 6.53 & 1.54 $\pm$ 1.38\\ 
 \hline 
  & Identified Core & 0.03 & 19.45 & 0.64 $\pm$ 0.89\\ 
Virial Ratio & Starless Core & 0.05 & 19.45 & 0.66 $\pm$ 0.91\\ 
  & Protostellar Core & 0.03 & 1.46 & 0.31 $\pm$ 0.30\\ 
 \hline 
\end{tabular}
  \label{tab:oriona_property}
\end{table}

\clearpage
\scriptsize
 \begin{longtable}[c]{cccd{-1}d{1}d{1}d{-2}d{-1}d{-1}d{1}d{-2}c}
 \caption[]{Properties of Identified Cores (1) \label{tab:catalog_position}}
  \\ \hline \hline 
 ID & RA & Dec & \multicolumn{1}{c}{$v_\mathrm{LSR}$} & \multicolumn{1}{c}{$A$}& \multicolumn{1}{c}{$2R_\mathrm{core}$} & \multicolumn{1}{c}{$2R_\mathrm{core}$}& \multicolumn{1}{c}{$a_\mathrm{major}$} & \multicolumn{1}{c}{$a_\mathrm{minor}$}& \multicolumn{1}{c}{PA} & \multicolumn{1}{c}{$dV_\mathrm{core}$} & \multicolumn{1}{c}{YSO} \\ 
  & (J2000) & (J2000) & \multicolumn{1}{c}{(km s$^{-1}$)} & \multicolumn{1}{c}{(arcsec$^2$)}& \multicolumn{1}{c}{(arcsec)} & \multicolumn{1}{c}{(pc)}& \multicolumn{1}{c}{(arcsec)} & \multicolumn{1}{c}{(arcsec)}& \multicolumn{1}{c}{(deg)} & \multicolumn{1}{c}{(km s$^{-1}$)} &  \\ 
 \hline \\ \endfirsthead 
 \caption[]{Continued} \\ 
 \hline \hline 
 ID & RA & Dec & \multicolumn{1}{c}{$v_\mathrm{LSR}$} & \multicolumn{1}{c}{$A$}& \multicolumn{1}{c}{$2R_\mathrm{core}$} & \multicolumn{1}{c}{$2R_\mathrm{core}$}& \multicolumn{1}{c}{$a_\mathrm{major}$} & \multicolumn{1}{c}{$a_\mathrm{minor}$}& \multicolumn{1}{c}{PA} & \multicolumn{1}{c}{$dV_\mathrm{core}$} & \multicolumn{1}{c}{YSO} \\ 
  & (J2000) & (J2000) & \multicolumn{1}{c}{(km s$^{-1}$)} & \multicolumn{1}{c}{(arcsec$^2$)}& \multicolumn{1}{c}{(arcsec)} & \multicolumn{1}{c}{(pc)}& \multicolumn{1}{c}{(arcsec)} & \multicolumn{1}{c}{(arcsec)}& \multicolumn{1}{c}{(deg)} & \multicolumn{1}{c}{(km s$^{-1}$)} &  \\ 
 \hline \\ \endhead 
 1 & 5 33 44.7 & -5 18 5 & 9.3 & 5738 & 85 & 0.17 & 21.8 & 19.3 & -149 & 0.38 & N  \\ 
 2 & 5 33 45.0 & -5 24 24 & 8.2 & 2756 & 59 & 0.12 & 25.0 & 10.8 & 89 & 0.18 & N  \\ 
 3 & 5 33 47.3 & -5 19 48 & 8.3 & 2925 & 61 & 0.12 & 30.8 & 8.9 & 84 & 0.17 & N  \\ 
 4 & 5 33 48.2 & -5 29 20 & 7.9 & 5119 & 81 & 0.16 & 34.1 & 9.1 & 61 & 0.28 & N  \\ 
 5 & 5 33 48.7 & -5 45 29 & 7.7 & 2925 & 61 & 0.12 & 20.5 & 13.6 & 74 & 0.22 & N  \\ 
 6 & 5 33 51.5 & -5 35 7 & 8.0 & 8100 & 102 & 0.20 & 24.3 & 20.3 & -175 & 0.34 & N  \\ 
 7 & 5 33 51.9 & -5 10 15 & 10.5 & 2756 & 59 & 0.12 & 16.2 & 11.1 & 80 & 0.25 & N  \\ 
 8 & 5 33 52.2 & -5 17 8 & 10.2 & 3263 & 64 & 0.13 & 24.3 & 12.5 & -138 & 0.19 & N  \\
\hline 

 \end{longtable}
\normalsize

\clearpage
\scriptsize
 \begin{longtable}[c]{cd{-1}d{-1}d{-1}d{-1}d{-2}d{-1}d{-1}d{-1}d{-1}}
 \caption[]{Properties of Identified Cores (2) \label{tab:catalog_property}}
  \\ \hline \hline 
 ID  & \multicolumn{1}{c}{$T_\mathrm{mb}^\mathrm{peak}$} & \multicolumn{1}{c}{$T_\mathrm{mb}^\mathrm{total}$}& \multicolumn{1}{c}{$M_\mathrm{core}$} & \multicolumn{1}{c}{$M_\mathrm{vir}$/$M_\mathrm{core}$}& \multicolumn{1}{c}{$\overline{n}$} & \multicolumn{1}{c}{$\overline{T}_\mathrm{ex}$}& \multicolumn{1}{c}{$\overline{N}_\mathrm{C^{18}O}$} & \multicolumn{1}{c}{$\overline{X}_\mathrm{C^{18}O}$} \\ 
  & \multicolumn{1}{c}{($\mathrm{K}$)} & \multicolumn{1}{c}{($\mathrm{K}$)} & \multicolumn{1}{c}{($M_\odot$)} &  & \multicolumn{1}{c}{(10$^4$ cm$^{-3}$)} & \multicolumn{1}{c}{($\mathrm{K}$)} & \multicolumn{1}{c}{(10$^{15}$ cm$^{-2}$)} &\multicolumn{1}{c}{(10$^{-7}$)}  \\ 
 \hline \\ \endfirsthead 
 \caption[]{Continued} \\ 
 \hline \hline 
 ID  & \multicolumn{1}{c}{$T_\mathrm{mb}^\mathrm{peak}$} & \multicolumn{1}{c}{$T_\mathrm{mb}^\mathrm{total}$}& \multicolumn{1}{c}{$M_\mathrm{core}$} & \multicolumn{1}{c}{$M_\mathrm{vir}$/$M_\mathrm{core}$}& \multicolumn{1}{c}{$\overline{n}$} & \multicolumn{1}{c}{$\overline{T}_\mathrm{ex}$}& \multicolumn{1}{c}{$\overline{N}_\mathrm{C^{18}O}$} & \multicolumn{1}{c}{$\overline{X}_\mathrm{C^{18}O}$} \\ 
  & \multicolumn{1}{c}{($\mathrm{K}$)} & \multicolumn{1}{c}{($\mathrm{K}$)} & \multicolumn{1}{c}{($M_\odot$)} &  & \multicolumn{1}{c}{(10$^4$ cm$^{-3}$)} & \multicolumn{1}{c}{($\mathrm{K}$)} & \multicolumn{1}{c}{(10$^{15}$ cm$^{-2}$)} &\multicolumn{1}{c}{(10$^{-7}$)}  \\ 
 \hline \\ \endhead 
 1 & 1.7 & 183.0 & 0.7 & 0.7 & 0.5 & 20.3 & 0.5 & 2.8  \\ 
 2 & 2.9 & 154.2 & 0.2 & 0.5 & 0.3 & 21.9 & 1.5 & 7.5  \\ 
 3 & 1.5 & 84.4 & 0.2 & 0.3 & 0.4 & 21.9 & 1.0 & 4.9  \\ 
 4 & 3.5 & 522.1 & 0.6 & 0.4 & 0.5 & 23.5 & 2.7 & 7.1  \\ 
 5 & 1.4 & 66.9 & 0.3 & 0.4 & 0.5 & 14.1 & 0.9 & 7.8  \\ 
 6 & 2.9 & 687.8 & 1.7 & 0.3 & 0.7 & 23.6 & 1.5 & 3.6  \\ 
 7 & 1.9 & 115.6 & 0.3 & 0.5 & 0.5 & 25.2 & 1.4 & 3.6  \\ 
 8 & 1.6 & 89.0 & 0.6 & 0.1 & 0.9 & 28.8 & 1.3 & 5.2  \\
\hline 

 \end{longtable}
\normalsize


\end{document}